\documentclass{aa}
\usepackage{graphicx}
\usepackage{txfonts}

\usepackage{color}
\usepackage{array}
\usepackage{caption}
\usepackage{subcaption}
\usepackage{amsmath,amssymb,amsfonts}
\usepackage{gensymb, footnote}
\usepackage{sidecap}
\usepackage{changepage}
\usepackage{tabu}
\usepackage[toc,page]{appendix}
\usepackage{multirow,multicol}
\usepackage[]{hyperref}
\hypersetup{
   colorlinks = true,
   citecolor ={cyan}
}

\usepackage{tablefootnote}
\usepackage{threeparttable}
\usepackage{upgreek}

\usepackage{stfloats}
\usepackage{placeins}

\newenvironment{Figure}
  {\par\medskip\noindent\minipage{\linewidth}}
  {\endminipage\par\medskip}

\date{March 2023}

\newcommand{\kms}{km\,s$^{-1}$}

\defcitealias{Bodensteiner_et_al_2020_Be}{Becat}

\begin{document}
   \title{HOney-BeeS}
   \subtitle{II. Be-X-ray binaries as testbeds for spectroscopic studies of Be stars} \titlerunning{Honey-Bees II. Be-X-ray binaries}

   \author{S. Janssens \inst{\ref{inst:kul},\ref{inst:tokyo}} 
           \and H. Sana\inst{\ref{inst:kul},\ref{inst:lgi}}
           \and T. Shenar\inst{\ref{inst:TAU}}
           \and J. Bodensteiner\inst{\ref{inst:api}}
           \and M. Abdul-Masih\inst{\ref{inst:IAC},\ref{inst:LaLag}}
          }
    \institute{
    {Institute of Astronomy, KU Leuven, Celestijnenlaan 200D, 3001 Leuven, Belgium\label{inst:kul}}
    \and   
    Research Center for the Early Universe, Graduate School of Science, University of Tokyo, Bunkyo, Tokyo 113-0033, Japan\\ \label{inst:tokyo}
    \email{soetkinjanssens@g.ecc.u-tokyo.ac.jp}
    \and
{Leuven Gravity Institute, KU Leuven, Celestijnenlaan 200D, box 2415, 3001 Leuven, Belgium \label{inst:lgi}}
    \and Tel Aviv University, The School of Physics and Astronomy, Tel Aviv 6997801, Israel \label{inst:TAU}
    \and
    Anton Pannekoek Institute for Astronomy, Science Park 904, 1098 XH, Amsterdam, The Netherlands \label{inst:api}    
    \and Instituto de Astrofísica de Canarias, C. Vía Láctea, s/n, 38205 La Laguna, Santa Cruz de Tenerife, Spain \label{inst:IAC}
    \and Universidad de La Laguna, Departamento de Astrofísica, Av. Astrofísico Francisco Sánchez s/n, 38206 La Laguna, Tenerife, Spain \label{inst:LaLag}
          }    

   \date{Received 10 August 2025 / Accepted 10 December 2025}

 
  \abstract
   {
   The majority of massive classical Be stars -- rapidly rotating stars with decretion discs -- are suggested to be binary interactions products. Their rapid rotation and often strong, variable, and emission-line dominated spectrum, make spectroscopic analysis challenging. Therefore, robust binary properties and meaningful statistical constraints are currently still lacking for the Be population.
   }
   {
   We aim to use Be-X-ray binaries, and their orbital periods derived from the X-rays, as benchmarks to investigate the reliability of different spectral lines and numerical methods for the measuring of radial-velocities and orbital period determination of Be stars.
   }
   {
   We use multi-epoch high-resolution HERMES spectra of a sample of seven Be-X-ray binaries and compare three different methods to determine radial velocities: cross-correlation, line-profile fitting, and the bisector method, on both absorption and emission lines. We also investigate the effect of different implementations of these methods.
   }
   {
   We find that cross-correlation is the most autonomous and convenient method. The bisector method is highly influenced by line-profile variability, while line-profile fitting requires a model template that does not encompass the complexity of Be star line profiles. Cross-correlation can also be implemented on non-Gaussian spectral lines and line blends seen in Be-star spectra. The statistical uncertainties on the radial-velocity precision that is reached by using cross-correlation on the high-resolution HERMES spectra is around 0.2-0.3\,km\,s$^{-1}$ for H$\alpha$ (emission) and $\sim 5$\,km\,s$^{-1}$ for absorption lines, like He {\sc i}, owing to the lower contrast between the line-depth and the noise level. Potential systematics are not included in the listed values.
   }
   { We recommend using cross-correlation as the standard method to determine radial velocities for Be stars as it is not dependent on any models and is easily compatible with line blends caused by the rapid rotation of the Be star. In general, whether the goal of a study is to analyse binary statistics of a population or perform an in-depth study of a specific system, we suggest using emission lines since they have a higher precision and less scatter than absorption lines. Here, H$\beta$ is preferred over H$\alpha$ because of its lesser variability. However, large-scale variability may cause large shifts in emission-line radial velocities, resulting in spurious eccentricities. In this case, orbital solutions should ideally be compared to the lower-signal absorption lines (if present). Finally, we highlight the need for understanding how companion-disc interactions alter the spectroscopic appearance of emission lines. 
   }


   \maketitle

\section{Introduction}
Classical Be stars are rapidly rotating B-type stars whose spectra show characteristic emission lines attributed to a decretion disc \citep{Struve_1931,Rivinius_2013}. About 20\%\ of B stars are of Be spectral type \citep{Porter_Rivinius_2003} and it is suggested that the Be phenomenon originates from single or binary evolution. Several independent studies have recently suggested that the majority of massive Be stars have a binary origin \citep{Boubert_Evans_2018, Bodensteiner_et_al_2020_Be, Hastings_et_al_2021, Dallas_et_al_2022, Dufton_et_al_2022, Dodd_et_al_2024,Klement_et_al_2024}. However, determining the binary statistics of the Be population remains challenging due to their discs, inducing (strong) variability of their emission lines, and to rapid rotation, which broadens their absorption lines significantly.\\
Several optical spectroscopic studies have successfully analysed Be binaries. Some studies focused on both emission and absorption lines \citep[e.g.][]{Poeckert_1981,Bjorkman_et_al_2002,Peters_et_al_2008,Naze_et_al_2022_gamCas,Janssens_et_al_2023}, while others focused more on emission lines \citep[e.g.][]{Koubsky_et_al_2012}. In some systems, the top and wings of the H$\alpha$ emission line may appear to move in anti-phase \citep[e.g. \object{LB-1};][]{Liu_et_al_2019_LB1, Abdul-Masih_et_al_2020,Shenar_et_al_2020} or the H$\alpha$ emission line appears rather stationary while there is actually motion of the star itself \citep[e.g. \object{HR6819};][]{Rivinius_et_al_2020_BH,Bodensteiner_et_al_2020_HR6819}. In both cases, black holes (BHs) were falsely assumed to reside in the system, suggesting that (H$\alpha$) emission lines could be non-optimal to study individual binaries.\\
The complexity and variability of the often double-peaked Be emission lines has different origins. One is the so-called violet-to-red (V/R) variability, where the strength of the blue and red emission peaks cyclically alternates \citep[][]{Okazaki_1991,Stefl_et_al_2009}. These variations are often attributed to a spiral density structure in the disc, which could be caused by a companion interacting with the disc \citep[][]{Panoglou_et_al_2016,Panoglou_et_al_2018_binarydiscs,Rubio_et_al_2025}. Moreover, variations in the top region of the emission lines from double peaked to triple or single peaked are also observed and mostly affect H$\alpha$. This variability can be attributed to a companion moving through the outer parts of the disc \citep[e.g.][]{Zamanov_et_al_2022}. Furthermore, the emission-line strength can cyclically vary over time and can, in some cases, even disappear. This is likely attributed to the formation and depletion of the disc itself \citep[e.g.][]{doazan_et_al_1986,Wisniewski_et_al_2010}.\\
So far, the use of photospheric absorption lines is generally preferred over the (variable and complex) emission lines. However, it is not always possible to resort to absorption lines for a spectroscopic analysis as many of them can be contaminated by emission \citep[e.g. several examples in this work and in][]{Koubsky_et_al_2012,Chojnowski_et_al_2018}. Moreover, because of their fast rotation, the signal in the broadened absorption lines is often spread and weakened as compared to emission lines, resulting in larger uncertainties and more scatter when determining radial-velocities (RVs). Finally, pulsational variability is also a major source of error for determining RVs of Be stars \citep[e.g.][]{Baade_1997,Baade_Balona_1994}. With the aforementioned challenges in mind, the spectroscopic analysis of Be stars could benefit from a standardised methodology to determine reliable RVs.\\
\\
Be-X-ray binaries (BeXRBs) are a subclass of the high-mass X-ray binaries (HMXBs), where the luminous companion is a Be star. Several of these BeXRBs show regular X-ray outbursts, which are often attributed to the passing of the compact companion through the dense disc. Hence, in many cases, orbital periods can be determined from the X-rays \citep[e.g.][]{Bongiorno_et_al_2011, Usui_et_al_2012_VES, Ferrigno_et_al_2013_VES, Moritani_et_al_2018}.\\
This paper is part of the Hermes Observational survey of OeBe Stars (Honey-BeeS). The Honey-BeeS is a multi-epoch spectroscopic program, in which over 200 bright (V $<$ 12 mag) northern Be stars are observed (see Bodensteiner, J. et al. in prep. for more details). The stars in Honey-BeeS are classical Be stars with spectral types earlier than B1.5 based on the BeSS database\footnote{\url{http://basebe.obspm.fr/basebe/}}. Given the spectral-type cut, the Be stars and their companions are both expected to be massive stars -- and thus compact-object progenitors --, such that we focus on double compact-object binary progenitors \cite[see][]{Bodensteiner_et_al_2020_Be}.\\
In this paper, we focus on the BeXRBs in HOney-BeeS. We aim to compare RVs obtained with different methods and different diagnostic lines (absorption and emission) to the orbital information that is found from the X-ray data. This way, we can investigate which methods and diagnostic lines are more reliable and provide the most precise results. \\
In Sect. \ref{sec_targets}, we describe the selection of the sample and the newly obtained spectroscopic data. The different methods to measure the RVs are explained in Sect. \ref{sec_methodology_RVs} including an overview of the mechanisms that are expected to give rise to RV variations. The short-term stability of different spectral lines and methods is reported in Sect.~\ref{sec_benchmarking_stability}. We compare different spectral lines and obtain orbital information on targets in Sect.~\ref{sec_periods_solutions}. We discuss our results in Sect.~\ref{sec_Discussion} and our conclusions in Sect.~\ref{sec_Conclusion}.

\section{Sample selection and observations}\label{sec_targets}
\subsection{Sample selection}
We searched for BeXRBs in HOney-BeeS through a cross-match with the X-ray catalogues of \citet{Liu_et_al_2000}, \citet{Reig_2011}, \citet{Walter_et_al_2015}, and \citet{Fortin_et_al_2023}. The cross-match resulted in eight matches, of which five were included in this study: \object{HD 259440}, \object{V420 Aur}, \object{V615 Cas}, \object{V725 Tau}, and \object{V831 Cas}. Both \object{MWC 656} \citep[][]{Janssens_et_al_2023} and \object{BD+47 3129} \citep{Naze_et_al_2022_gamCas} have already been analysed by means of different methods and spectral lines and were excluded. The system \object{2E 1752} was not yet analysed in this work due to a lack of data.
\\
We included two additional targets that were not listed in BeSS at the time of making the HOney-BeeS catalogue: \object{EM* VES 826} and \object{X Per}. 
In total, we thus have seven BeXRB targets, for which a short description and spectral overview is given in appendix \ref{appendix_target_summary}. The targets and their basic information are listed in Table~\ref{table_targets_general_info}.

\subsection{Observations}
We used the High-Efficiency and high-Resolution Mercator Echelle Spectrograph (HERMES), mounted on the 1.2m Mercator telescope at the Roque de los Muchachos observatory in La Palma, Spain \citep{Raskin_2011}. The data were collected with the high-resolution mode, delivering a spectral resolving power of $R = \lambda/\Delta\lambda \simeq 85 000$ and covering the wavelength range 3800-9000 $\AA$ with a step size of $\sim$0.02-0.05 $\AA$. Most of the observations were randomly spread between 2020 and 2024, within the frame of the HOney-BeeS program (see Bodensteiner, J. et al. in prep.). Additionally, all archival HERMES spectra past a two-year proprietary period were also used in this work.\\
Raw spectra were reduced with the HERMES reduction pipeline (HERMES-DRS, V .7.0), which includes bias correction, flat-fielding, wavelength calibration, cosmic and dark removal, and correction for barycentric motion. All data were normalised with a spline interpolation using the Python package {\tt SciPy} \citep{Virtanen_2020}. 
The typical signal-to-noise (S/N) per pixel at $\sim$4500 $\AA$, and number of spectra $N_{\text{spec}}$ are listed in Table \ref{table_targets_general_info}. The time baseline of the observations is also given in Appendix \ref{appendix_target_summary}.

\begin{table*}
    \centering
\caption{Target names, spectral type, X-ray orbital period $P_{\text{X}}$, eccentricity $e$, spin period $P_{\text{spin}}$, V-band magnitude, S/N around 4500$\AA$, and number of available spectra $N_{\text{spec}}$.}
    \begin{tabular}{ c|cccccccc }
     \hline
     \hline
    Target id & spectral type & $P_{\text{X}}$ [d] & $e$ & $P_{\text{spin}}$ [s] & V mag & S/N at 4500$\AA$ & $N_{\text{spec}}$ \\ 
     \hline
      \object{EM* VES 826} & B0.2\,Ve & 150-155 & /$^{\text{c}}$ & 205 & 10.73 & 20 &9\\
      \object{HD 259440} & B0pe & $\sim$ 317 & ?$^{\text{d}}$ & /$^{\text{c}}$ & 9.12&70&38\\
      \object{V420 Aur} & B0\,IVpe & /$^{\text{c}}$ & /$^{\text{c}}$ & /$^{\text{c}}$ &7.48 &35/70$^{\text{a}}$& 187\\
      \object{V615 Cas} & B0\,Ve & 26.5 & ?$^{\text{d}}$ & 0.269 & 10.75 &35/50$^{\text{b}}$&47\\
      \object{V725 Tau} & O9.5\,IIIe & 110 & 0.47 & 104 & 9.39 &70&29\\
      \object{V831 Cas} & B1\,IIIe & /$^{\text{c}}$ & /$^{\text{c}}$ & 1400 & 11.37 & 30 & 15\\
      \object{X Per} & B0/1\,Ve & 250 & 0.11 & 837 & 6.72 &120&20\\
     \hline
    \end{tabular}
         \flushleft
    \begin{tablenotes}
      \small \textbf{Notes} See Appendix \ref{appendix_target_summary} for references on periods. $^{\text{a}}$\,The first 142 spectra have S/N $\sim$ 35, while the others have S/N $\sim$ 70. $^{\text{b}}$\,The first 18 spectra have S/N $\sim$ 35, while the others have S/N $\sim$ 70. $^{\text{c}}$\,Unknown. $^{\text{d}}$\,Multiple values reported in literature.
    \end{tablenotes}
\label{table_targets_general_info}
\end{table*}

\section{Methods for radial velocity measurements}\label{sec_methodology_RVs}
We used three methods to obtain RVs: cross-correlation (CC, Sect. \ref{sec_cross_correlation_method}), the bisector method (labeled `bisec' in figures, Sect. \ref{sec_bisector_method}), and line-profile fitting (lpf, Sect. \ref{sec_gaussian_fitting_method}). In Sect. \ref{sec_intrinsic_variability}, we discuss intrinsic variabilities that can affect the RV measurements.

\subsection{Cross correlation}\label{sec_cross_correlation_method}
By cross-correlating an observational template or a model spectrum with the individual spectra, a CC function is obtained. By fitting a parabola or Gaussian to the maximum region of the CC function, RVs are determined \citep[][]{Zucker_2003}. Statistical errors are calculated as described in \citet{Zucker_2003}. \\
In this work, CC was done iteratively. In the first iteration, the first spectrum was used as a template. In further iterations, the used template was an average spectrum according to the RVs obtained in the previous iteration. The higher S/N in the templates of later iterations typically lowers the RV uncertainties, especially for weaker lines. The procedure stopped when the RVs  of the last three iterations were in agreement within 0.01\,\kms. \\
The derived RVs are not absolute but relative to the master template. Thus, no systemic velocity can be determined. For the goal of this work, which is determining if orbital motion can be detected in BeXRBs, this poses no problem.

\subsection{Bisector}\label{sec_bisector_method}
To obtain the bisector of a given spectral line, the line is first divided into bins of identical flux ranges in the red and blue wing. In each blue and red flux bin, intermediate RVs are calculated by averaging the RVs of each wavelength point in the bin with respect to the rest wavelength of the spectral line, which were taken from the National Institute of Standards and Technology (NIST) spectral line database\footnote{\url{https://pml.nist.gov/PhysRefData/ASD/levels_form.html}}. The intermediate RVs of two identical flux bins in the red and blue wing are then averaged to obtain the mean RV ($\overline{\text{RV}}$) of a given flux bin. The $\overline{\text{RV}}$s trace the bisector of the spectral line. The final RV of a spectral line is then obtained by averaging all $\overline{\text{RV}}$s and its uncertainty is determined by the standard deviation of the $\overline{\text{RV}}$s. The derived uncertainty on the RV of the spectral line thus depends on how symmetric the line shape is and will be, in general, very small. \\
The bisector is sensitive to asymmetries in the line and line-profile changes can easily be traced through the bisector. Therefore, it is often used to trace variability in the line profile, but it has already been used to determine RVs \citep[e.g.][]{bisector}. We investigate the effect of excluding different bottom and top regions of the lines on the RV determination in Appendix \ref{appendix_boundaries_methods}.

\subsection{Line-profile fitting}\label{sec_gaussian_fitting_method}
We fitted Gaussian or Voigt profiles to the spectral lines, with respect to the NIST rest wavelength, by using the python package {\tt lmfit} \citep{lmfit}. We used several fit setups: (i) the full spectral line with a single emission or absorption profile, (ii) part of the line while excluding the (variable) top region of an emission line, or (iii) a combination of an emission and absorption profile, which is only fitted for double-peaked emission lines. A cut-off top fit and an emission + absorption fit are marked with `cut' and `double', respectively, in figures.\\ 
For double fits, the central wavelengths of the emission and absorption profile were set to be identical, such that they share the same RV. Should we not enforce this, V/R variations in the emission profiles will lead to fits displaying non-equal central wavelengths for the fitted profiles and apparent anti-phase motions.

\subsection{Uncertainties and intrinsic variability}\label{sec_intrinsic_variability}
One of our goals is to investigate how line variability affects the accuracy of RV measurements obtained from various methods. Informed by these results, we aim to identify variability criteria that would separate intrinsic line-profile variations (e.g. due to discs) from those induced by a companion in a binary system. In this context, we adapt the formalism of \citet{Dsilva_et_al_2020} for Wolf-Rayet star RV measurements. We consider four main sources of variability in our RV measurements: measurement uncertainties, disc variability, binarity, and pulsations, such that 
\begin{equation}
\text{RV}_{\text{meas}} = \text{RV}_{\text{true}} + \theta_{\text{meas}} + \theta_{\text{disc}} + \theta_{\text{bin}} + \theta_{\text{puls}},
\end{equation}
where $\text{RV}_{\text{meas}}$ is the measured RV and $\text{RV}_{\text{true}}$ is the true velocity of the star. The $\theta_{\text{meas}}$, $\theta_{\text{disc}}$, $\theta_{\text{bin}}$, and $\theta_{\text{puls}}$ are (potential) contributions of the measurement process, disc, orbital motion, and pulsations to the measured RV, respectively.\\
We can take advantage of the fact that these four quantities have a different statistical behavior: 1) $\theta_{\text{meas}}$ is expected to be mostly affected by random noise, likely Gaussian distributed, and impacted by quantities such as the S/N and the line width. 2) $\theta_{\text{disc}}$ will mostly impact emission lines, resulting from asymmetries in the disc arising on timescales of days to decades \citep[e.g.][]{Labadie-Bartz_et_al_2018}. These variabilities can be intrinsic or induced by a companion passing through the disc. Both might shift the RV measurements compared to the RV of the centre-of-mass of the star. 3) $\theta_{\text{bin}}$ is the orbital motion contribution of the Be star in a binary and will vary periodically over time following Kepler's law. Timescales for orbital periods are expected to be of the order of months to years if Be stars are post-interaction products \citep[e.g.][]{Hastings_et_al_2021,Schurmann_2025}. 4) $\theta_{\text{puls}}$ represents pulsationally induced RV variability, which might be more prominent for the absorption lines. This variability can have time spans of hours to days, and is quite diverse for Be stars \citep[e.g.][]{Porter_Rivinius_2003}.\\
Under the assumption that these quantities are uncorrelated, we can express the variance of our RV sample as 
\begin{equation}
\sigma_{\text{tot}}^2 = \sigma_{\text{meas}}^2 + \sigma_{\text{disc}}^2 + \sigma_{\text{bin}}^2 + \sigma_{\text{puls}}^2.
\end{equation}
On time scales of hours, the ideal case gives us $\sigma_{\text{tot}}^2 \approx \sigma_{\text{meas}}^2$. However, this may be different than the statistical error obtained from the method itself due to, for example, normalisation artifacts or pulsations ($\sigma_{\text{puls}}$). Nevertheless, this suggests that we can use short cadence RV measurement time series to evaluate $\sigma_{\text{meas}}$ and long timebase series of a single star to place constraints on $\sigma_{\text{disc}}$. Moreover, emission lines and absorption lines might not have an equal response to pulsational or disc variability. We discuss these aspects further in Sect. \ref{sec_benchmarking_stability}.\\
In reality $\sigma_{\text{disc}}$ and $\sigma_{\text{bin}}$ may not be fully uncorrelated, for example when the companion periodically passes through the disc, which is the case for some BeXRBs. Therefore, when measuring RVs from Be stars, a by-eye assessment of the variability in the spectra is necessary to evaluate if the measured RV changes can be true RV measurements for the star or whether they are inferred from disc variability.

\begin{figure*}
     \centering
    \begin{subfigure}{0.333\linewidth}
    \includegraphics[trim= 0 15 0 15,clip,width = \textwidth]{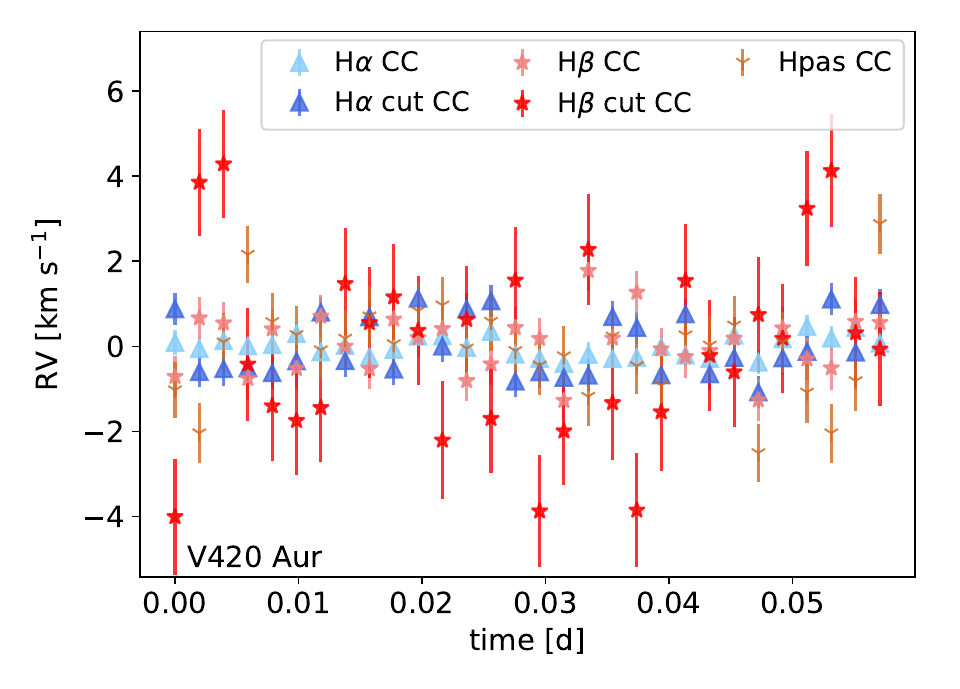}
    \end{subfigure}
    \hspace{-0.1in}
     \begin{subfigure}{0.333\linewidth}
    \includegraphics[trim= 0 15 0 15,clip,width = \textwidth]{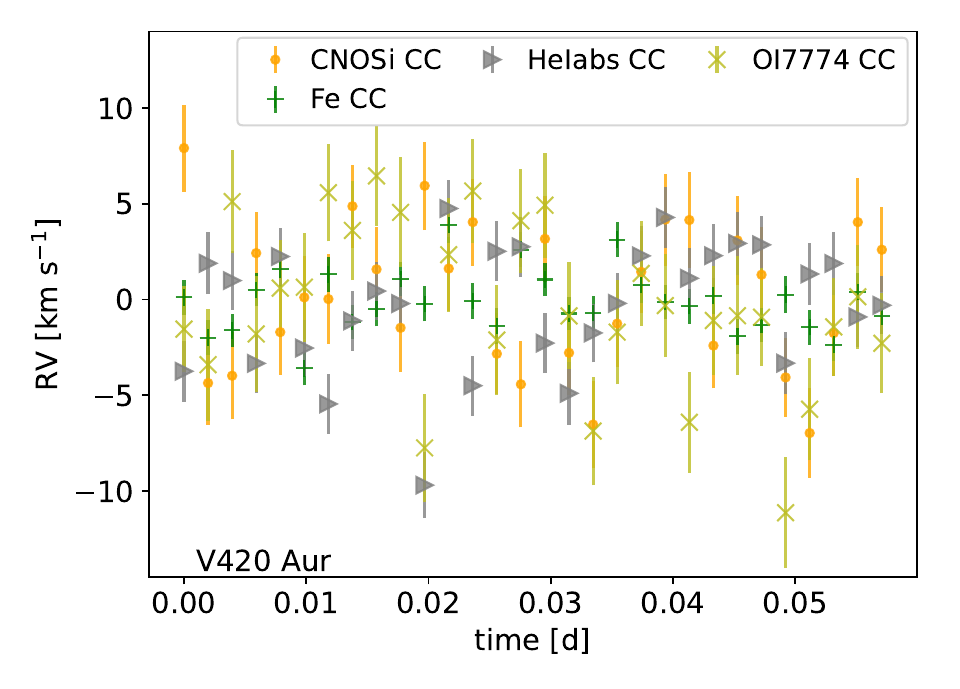}
    \end{subfigure}
    \hspace{-0.1in}
    \begin{subfigure}{0.333\linewidth}
    \includegraphics[trim= 0 15 0 15,clip,width = \textwidth]{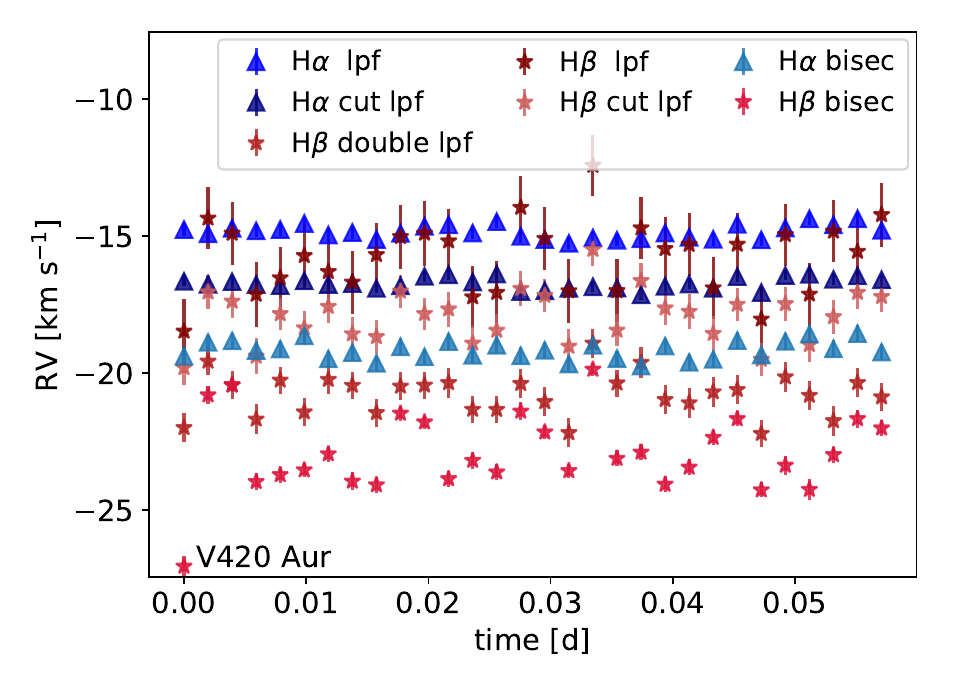}
    \end{subfigure}
    \begin{subfigure}{0.333\linewidth}
    \includegraphics[trim= 0 15 0 15,clip,width = \textwidth]{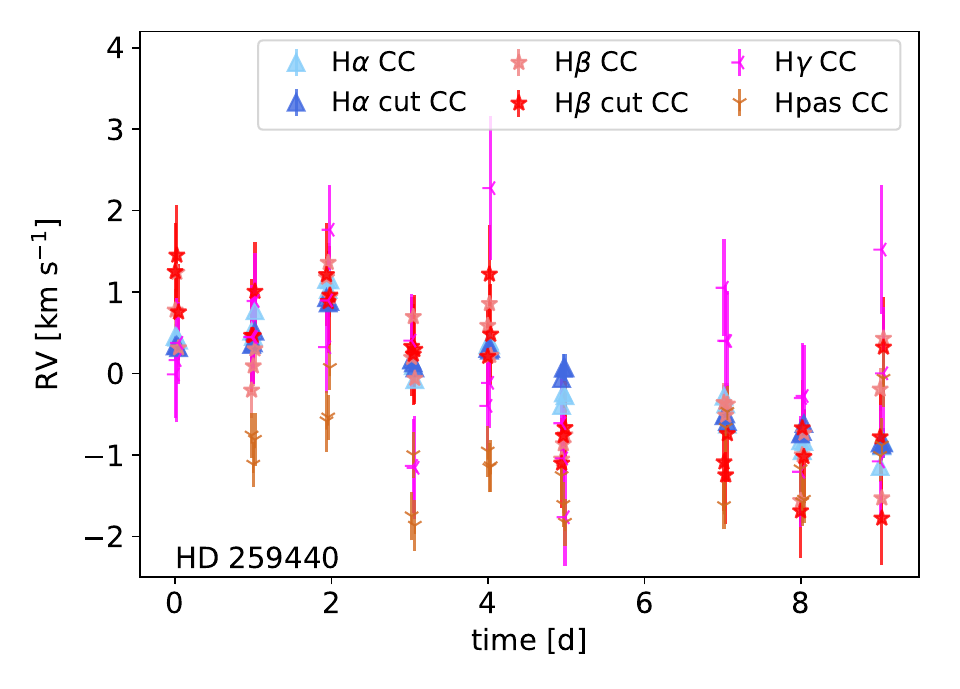}
    \end{subfigure}
    \hspace{-0.1in}
     \begin{subfigure}{0.333\linewidth}
    \includegraphics[trim= 0 15 0 15,clip,width = \textwidth]{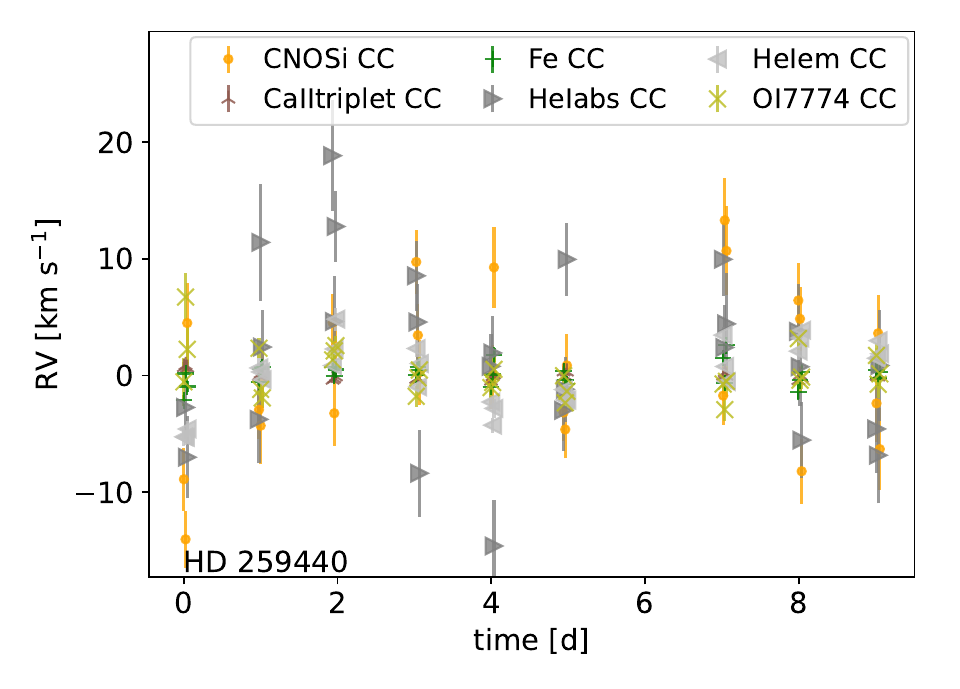}
    \end{subfigure}
    \hspace{-0.1in}
    \begin{subfigure}{0.333\linewidth}
    \includegraphics[trim= 0 15 0 15,clip,width = \textwidth]{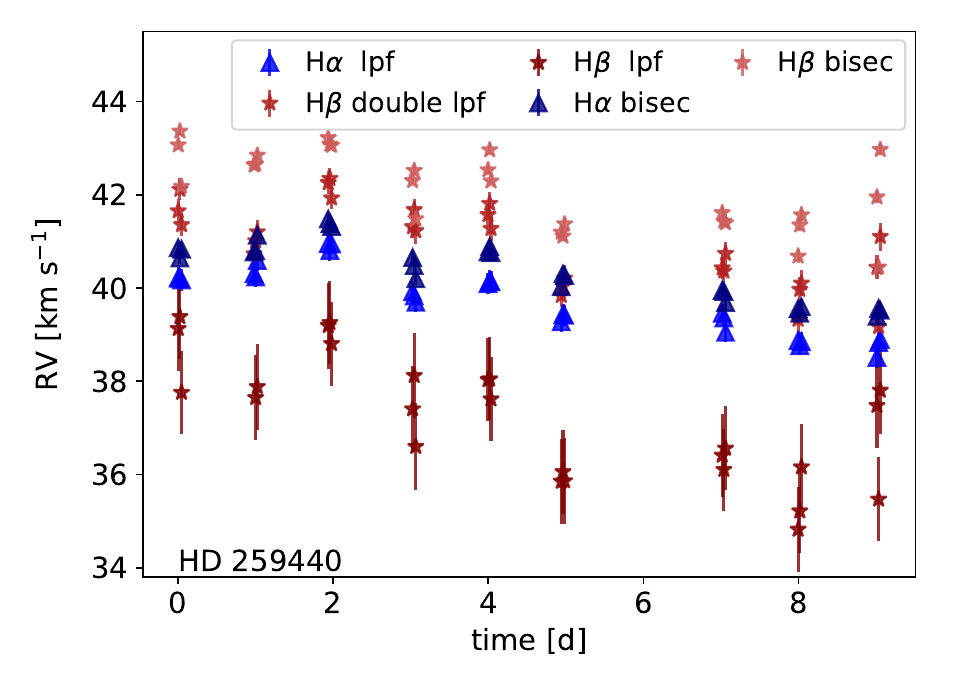}
    \end{subfigure}
    \begin{subfigure}{0.333\linewidth}
    \includegraphics[trim= 0 15 0 15,clip,width = \textwidth]{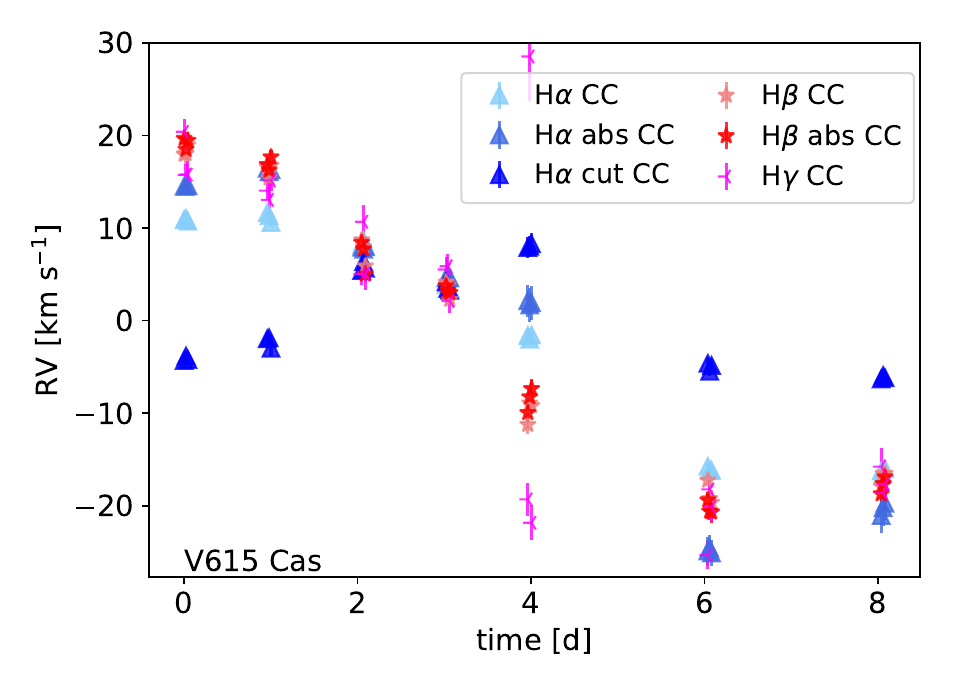}
    \end{subfigure}
    \hspace{-0.1in}
     \begin{subfigure}{0.333\linewidth}
    \includegraphics[trim= 0 15 0 15,clip,width = \textwidth]{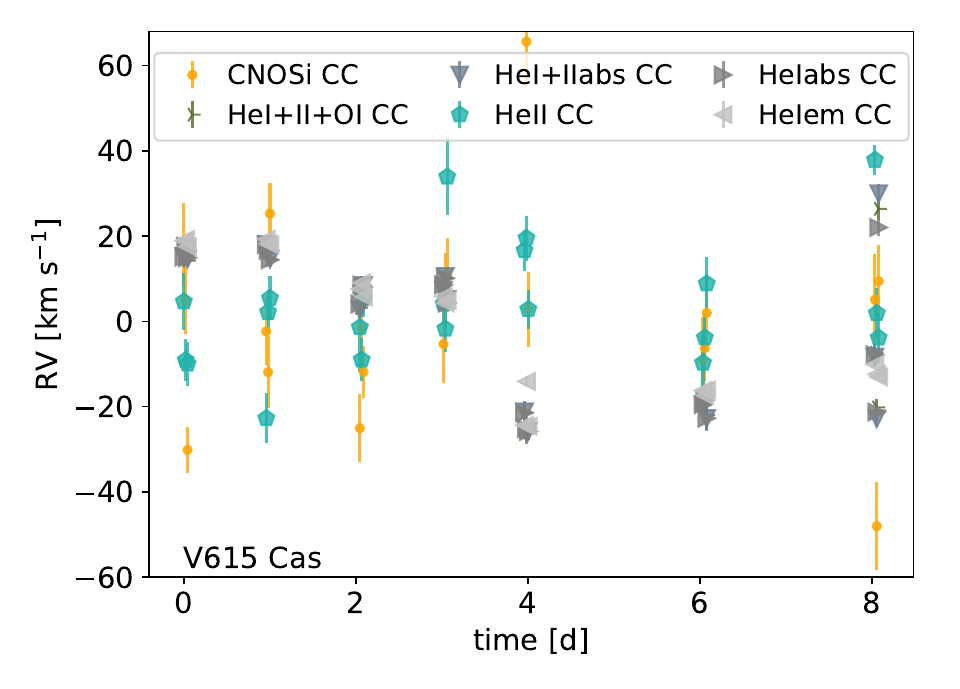}
    \end{subfigure}
    \hspace{-0.1in}
    \begin{subfigure}{0.333\linewidth}
    \includegraphics[trim= 0 15 0 15,clip,width = \textwidth]{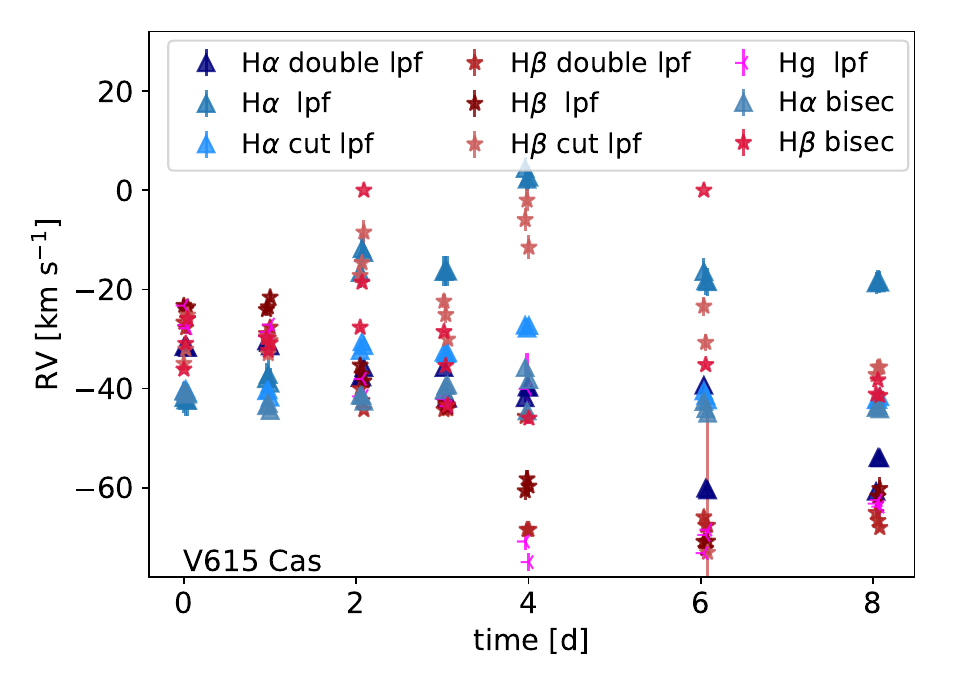}
    \end{subfigure}
    \caption{Radial velocities obtained with different methods on different spectral lines for \object{V420 Aur} (top row) for one night, \object{HD 259440} (middle row) for all nights with three consecutive spectra, and \object{V615 Cas} (bottom row) for all nights with three consecutive spectra. The x-axis shows the time spectra were taken since the first spectra shown. Left: hydrogen lines and CC. Middle: non-hydrogen lines and CC. Right: hydrogen lines and lpf and bisector. In the legend, `Fe' refers to the Fe {\sc ii} emission lines in the region of 5200$\AA$.
    }
    \label{fig_RVs_different_methods}
\end{figure*}

\section{Benchmarking methods and diagnostic lines}\label{sec_benchmarking_stability}
In this section, we determine the stability of different spectral lines and methods on short timescales, providing us with an estimate of $\sigma_{\text{meas}}$. Indeed, on short time scales of $\sim$hours, we expect no large RV variations, as suggested in Sect. \ref{sec_intrinsic_variability} (except for potential pulsations). For this purpose, we used targets for which high-cadence monitoring over a few hours time span is available, which are \object{V420 Aur}, \object{HD 259440}, and \object{V615 Cas}.\\ 
We assess which method and spectral line yield a better precision by calculating the root-mean-squared (rms), which is an estimator of the standard deviation using \citep{Bevington_Keith_stats}:
\begin{equation}\label{eq_RMS_bexrbs}
\text{rms} = \frac{N}{N-1}\sqrt{\sum_i \frac{(x_i - \bar{x})^2}{\sigma_i^2} / \sum_i \frac{1}{\sigma_i^2}},
\end{equation}
where $x_i$ is the $i$-th RV measurement with uncertainty $\sigma_i$, $\bar{x}$ is the average RV measurement over the considered time span, and $N$ is the number of RV data points. We describe the used lines and methods in Sect. \ref{subsec_methods_diagnostic_lines} and discuss the results in Sect. \ref{subsec_stability_summary}.

\subsection{Used methods and diagnostic lines}\label{subsec_methods_diagnostic_lines}
For V420 Aur, we used the CC method on H$\alpha$, H$\beta$, the H Paschen series, the Fe {\sc i} emission lines in the region of 5150-5380$\AA$ (listed as `Fe'), the He {\sc i} lines that are in absorption (those marked by a single `a' in Table \ref{table_spectral_lines_alltargets}, further referred to as `He {\sc i} abs'), the O\,{\sc i}\,$\lambda\lambda$7772,74,75 emission line complex (listed as `OI7774'), and the N {\sc ii}, Si {\sc iv}, N {\sc iii}, O {\sc ii}, and C {\sc iii} lines between 4630-4655$\AA$, which are blended due to the rapid rotation of the Be star. We refer to this line blend as the CNOSi line blend. For the bisector and lpf methods, only H$\alpha$ and H$\beta$ were used.\\
For \object{HD 259440}, we used the same set of lines as for \object{V420 Aur}. Additionally, we also measured RVs for H$\gamma$, the He {\sc i} emission lines (those marked with `e' or `dp' in Table \ref{table_spectral_lines_alltargets}, further referred to as `He {\sc i} em'), and the Ca {\sc ii}~$\lambda\lambda$8498, 8542, 8662 (further referred to as `Ca {\sc i} triplet'). For the Ca {\sc i} triplet, the separation from the H Paschen series is challenging and hence the fit is mostly focused on the blue wing of the Ca emission.\\
For V615 Cas, we used the CC method on H$\alpha$, H$\beta$, H$\gamma$, He {\sc i} abs, He {\sc i} em, the He {\sc ii} lines (in absorption), a combination of He {\sc i} abs and He {\sc ii} with and without O {\sc i} 7774, and the CNOSi blend. We focused the H$\alpha$ fit on the full line, the emission wings (`cut'), and the absorption component (`abs'). For the lpf and bisector method, we used H$\alpha$ and H$\beta$. We also fitted H$\gamma$ using lpf with an absorption profile. For the single-Gaussian fitted H$\beta$ (`H$\beta$ lpf'), the fit focused on the absorption core and not on the emission.

\begin{figure*}
    \sidecaption
     \centering
    \includegraphics[trim= 0 15 0 13,clip,width=\linewidth]{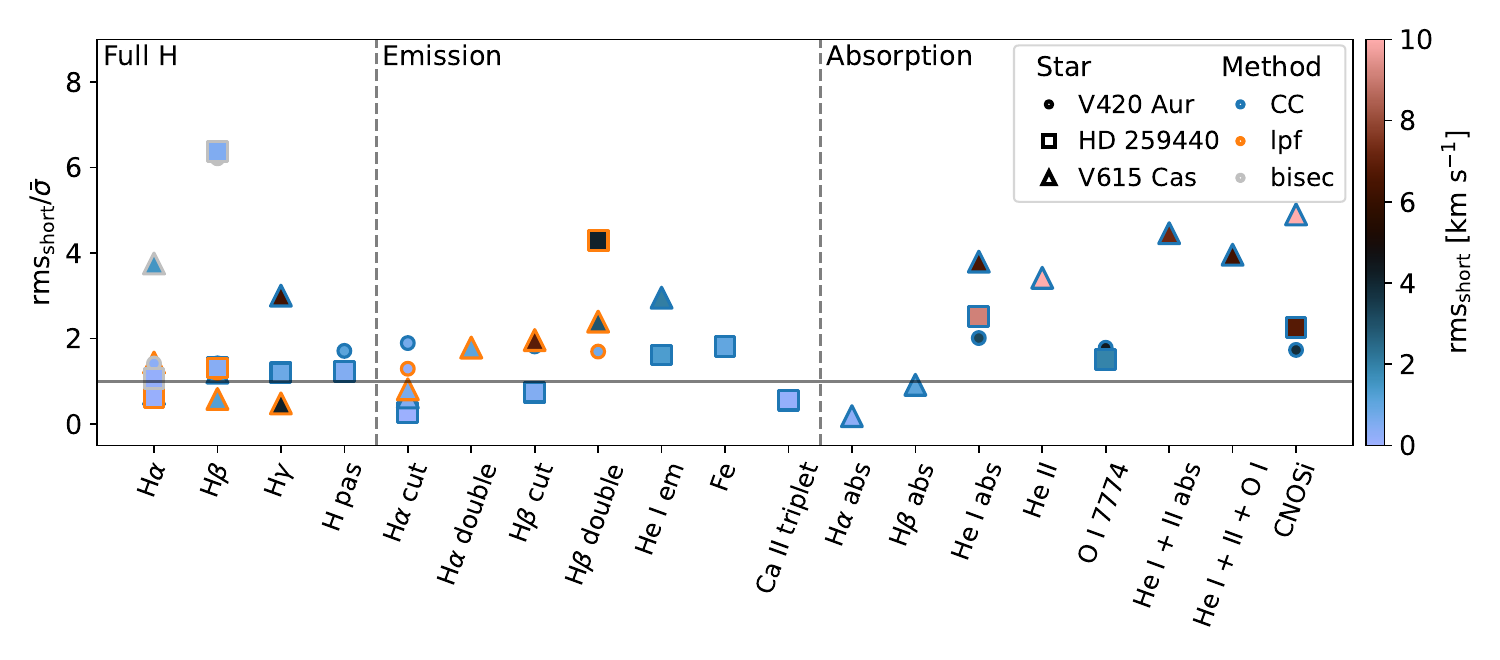}
    \caption{Ratio of $\mathrm{rms}_\mathrm{short}$ and $\bar{\sigma}$ for different spectral lines (indicated on the x-axis) for \object{V420 Aur} (circles), \object{HD 259440} (squares), and V615 Cas (triangles). Different methods are indicated with different edge colors. The color map in the symbols indicates the value of $\mathrm{rms}_\mathrm{short}$. A value of 1 for $\mathrm{rms}_\mathrm{short}/\bar{\sigma}$ is shown with a solid line.
    }
    \label{fig_RVs_line_precision}
\end{figure*}

\subsection{The stability of the methods and diagnostic lines}\label{subsec_stability_summary}
The measured RVs for \object{V420 Aur} are shown in the top panels of Fig. \ref{fig_RVs_different_methods} for the 30 consecutive spectra obtained within a time span of 2h30min in the night of  October 12, 2016. The RVs of \object{HD 259440} and \object{V615 Cas} are displayed in the middle and bottom rows of Fig. \ref{fig_RVs_different_methods}, respectively, showing the nights for which three consecutive spectra are available (November 2009). The ratio of the average measurement uncertainties ($\bar{\sigma}$) and rms values for the consecutive spectra ($\mathrm{rms}_\mathrm{short}$) are shown in Fig. \ref{fig_RVs_line_precision} and individual values are given in Table \ref{table_chi2values} for the three systems.\\
The rms short is larger than $\bar{\sigma}$ for almost all spectral lines except for H$\alpha$, which indicates that the measurement uncertainties $\sigma_{\text{meas}}$ obtained from the methods are underestimated. Hence, a smaller precision can be reached on the orbital analysis than what is given by the $\sigma_{\text{meas}}$. For example, $\sigma_{\text{meas}}$ for H$\beta$ with CC are underestimated by a factor 1.4-2, even increasing to a factor $>5$ when using the bisector method. This extreme increase in the underestimation of $\sigma_{\text{meas}}$ with the bisector method is because the method is much more sensitive to line-profile variability and the method's intrinsic $\sigma_{\text{meas}}$ are very small by construction.\\
The rms short of absorption lines is larger than for emission lines. For example, typical precisions reached using CC on H$\alpha$ are 0.2-0.3\,km\,s$^{-1}$, as compared to $\sim$5~km\,s$^{-1}$ for absorption lines. Generally, the contrast between the intensity of the absorption lines and the noise level is lower than for the emission lines, which could increase the scatter. However, this should already be accounted for in $\sigma_{\text{m}}$. Instead, there is a high possibility that the rms short increases because pulsational variability at the stellar surface is impacting more effectively the (photospheric) absorption lines than the (disc-dominated) emission lines. 
\\
From Figs. \ref{fig_RVs_different_methods} and \ref{fig_RVs_line_precision}, it is clear that stronger (emission) lines, like H$\alpha$ and H$\beta$, generally show less scatter than weaker (absorption) lines, such as He {\sc i} abs, O {\sc i} 7774, or the CNOSi blend. Based on this diagnostic, absorption lines should not necessarily be preferred over emission lines.\\
However, in the example of \object{V615 Cas}, H$\alpha$ shows significant differences compared to other lines. Depending on which part of the line is used (full line `H$\alpha$', wings only `H$\alpha$ cut', or absorption part only `H$\alpha$ abs'), the derived RVs differ significantly. While the H$\alpha$ absorption component follows the same trend as He {\sc i} lines and H$\beta$, the emission parts and the combined fit do not. This is most likely not seen in H$\beta$ because its emission component is rather weak and the absorption feature dominates. The same is seen for the `H$\alpha$ lpf' and `H$\alpha$ bisec' as both are tracing the emission-line RV.\\
\\
For lpf, fitting double-Gaussian profiles is often challenging as good initial guesses need to be provided. Otherwise, the fit can have trouble finding the correct absorption or emission feature if they are rather weak. Using a single Gaussian or Voigt profile often reduces the impact of the initial guesses. However, the fitting becomes less accurate for emission lines with strong absorption cores as the method tries to find a compromise between the emission and the absorption. In this case, cutting out the absorption feature from the single-Gaussian fit improves the fit quality.\\
\\
To conclude, overall, the precision with which RVs can be determined is of the order of a few km\,s$^{-1}$, even when using the weaker absorption lines. While all three methods can yield satisfactory results, CC and bisector are more autonomous than lpf, which often needs tweaking of initial guesses. However, the measurement uncertainties are largely underestimated with the bisector method. Moreover, CC can also be fitted on line blends (e.g. CNOSi) and can easily account for asymmetric line shapes, which is difficult with the bisector.\\
However, CC does not yield a systemic velocity. This is not a problem for science goals that are looking at relative variability only, but might be a limitation for other goals. Performing CC with a theoretical template may alleviate this issue. \\
The often weak signal in the absorption lines combined with the pulsational variability can result in a large scatter on the RVs and hence less accuracy. Based on the precision that is reached, emission-line RVs can be a good alternative. However, H$\alpha$-based RVs should be considered with caution as already shown in two BH imposters \citep[\object{LB-1} and \object{HR6819};][]{Liu_et_al_2019_LB1, Abdul-Masih_et_al_2020, Bodensteiner_et_al_2020_HR6819, Rivinius_et_al_2020_BH}. While H$\alpha$ is the strongest line in the optical spectrum, it is also the most variable and can lead to RVs that are not following the orbital motion (see also Sect. \ref{sec_periods_solutions}). We suggest that H$\beta$, and most likely higher-order Balmer lines, can be a good alternative to absorption lines for RV measurements in the spectra of Be stars. We will further test the reliability of individual spectral lines in Sect. \ref{sec_periods_solutions}, where we search for orbital solutions for the BeXRBs. 

\begin{table*}
    \centering
\caption{Orbital solutions for \object{V615 Cas} using different spectral lines. }
    \begin{tabular}{ cccccccc}
     \hline
     \hline
spectral line&$P$ [d]&$e$&$\omega$ [$\degree$]&T0&$K_1$ [km\,s$^{-1}$]&$z_0$ [km\,s$^{-1}$]&rms\\
\hline
H$\beta$&26.506\,$\pm$\,0.003&0.58\,$\pm$\,0.13&88\,$\pm$\,21&10494.39\,$\pm$\,0.68&14.1\,$\pm$\,2.6&$-$23.4\,$\pm$\,1.6&7.5\\
H$\alpha$ abs&26.509\,$\pm$\,0.003&0.53\,$\pm$\,0.10&61\,$\pm$\,14&10493.89\,$\pm$\,0.72&22.8\,$\pm$\,2.7&$-$28.0\,$\pm$\,1.9&11\\
H$\alpha$ em&26.508\,$\pm$\,0.008&0.69\,$\pm$\,0.18&265\,$\pm$\,18&10499.8\,$\pm$\,1.1&20.5\,$\pm$\,6.7&14.9\,$\pm$\,1.8&7.2\\
H$\gamma$&26.509\,$\pm$\,0.003&0.70\,$\pm$\,0.11&109\,$\pm$\,14&10495.41\,$\pm$\,0.64&21.3\,$\pm$\,3.3&$-$23.3\,$\pm$\,2.2&11\\
He {\sc i} abs&26.489\,$\pm$\,0.009&0.00\,$\pm$\,0.15&90&10506.02\,$\pm$\,0.82&19.8\,$\pm$\,3.1&$-$27.9\,$\pm$\,2.5&11\\
\hline  
H$\beta$&31.055\,$\pm$\,0.005&0.30\,$\pm$\,0.12&62\,$\pm$\,27&10356.9\,$\pm$\,1.9&15.3\,$\pm$\,2.3&$-$26.4\,$\pm$\,1.6&6.8\\
H$\alpha$&31.055\,$\pm$\,0.007&0.29\,$\pm$\,0.11&81\,$\pm$\,30&10358.2\,$\pm$\,2.2&25.2\,$\pm$\,4.3&$-$30.8\,$\pm$\,2.8&10\\
He {\sc i} abs &26.093\,$\pm$\,0.003& 0.57\,$\pm$\,0.08&348\,$\pm$\,13&10329.46\,$\pm$\,0.55&21.4\,$\pm$\,2.4&$-$29.6\,$\pm$\,1.8&9.9\\
\hline
\end{tabular}
         \flushleft
    \begin{tablenotes}
      \small \textbf{Notes} Listed uncertainties correspond to 1$\sigma$. The last three rows are solutions when using the periods found in the periodograms as initial guesses.
    \end{tablenotes}
\label{table_orbital_V615_Cas}
\end{table*}
\section{Searching for periodicities and orbital solutions}\label{sec_periods_solutions}
In this section, we will present the RV measurements of the targets, together with an orbital fit (if there is sufficient data). We only present results using CC and compare different spectral lines. For \object{V420 Aur}, \object{HD 259440}, and \object{V615 Cas}, we merged the consecutive spectra, taken on timescales of a few hours, to increase the S/N of the spectra for these nights.\\
Orbital fitting was performed using spinOS\footnote{\url{https://github.com/matthiasfabry/spinOS}} \citep{Fabry_et_al_2021}, which uses a non-linear least-squares minimisation. The fitted parameters are orbital period $P$, eccentricity $e$, argument of periastron $\omega$\footnote{spinOS fits $\omega$ with respect to the companion. Here we list $\omega$ with respect to the Be star (i.e. a $180\degree$ shift from the rteurned spinOS value).}, time of periastron passage T0 (given in BJD $-$ 2\,450\,000), semi-amplitude of the RV curve $K_1$, and the systemic velocity. Since the RVs are measured using CC, this systemic velocity has no physical meaning and is referred to as the zero-point offset $z_0$.\\
For all targets, we also calculated periodograms (for details, see Appendix \ref{appendix_extraPs}). For none of the targets is the X-ray orbital period the most significant period. This already highlights the difficulty of obtaining independent orbital periods and solutions.\\
\\
We divide our sample into three groups, based on the variability of the emission lines and the prominence of the absorption lines (see Appendix \ref{appendix_target_summary} for an overview of the spectral appearance of the targets). First, we have stars with strong absorption lines and clearly double-peaked emission lines without large-scale variability in our time-series spectra (`group A', Sect. \ref{sec_groupA}), in which we only put V615 Cas. Second, we have stars where emission lines are also stable, as for group A, but the signal in absorption lines is rather weak as they are very broadened, and emission lines are not necessarily double peaked (`group S', Sect. \ref{sec_groupS}). Here, we find EM* VES 826 and HD 259440. Third are objects which show large-scale variability in their emission lines (`group LV', Sect. \ref{sec_groupLV}), which contains V725 Tau, V831 Cas, and X Per.\\
Lastly, we did not yet classify V420 Aur. For years, V420 Aur shows rather stable absorption lines. Only in a few recent spectra is there notion of V/R variability as well as an apparent increase in emission-line strength. Therefore, we consider V420 Aur to be of group LV, but discuss it separately.

\begin{figure*}
    \centering
    \begin{subfigure}{0.49\linewidth}
    \includegraphics[trim = 5 15 10 15, clip, width = \textwidth]{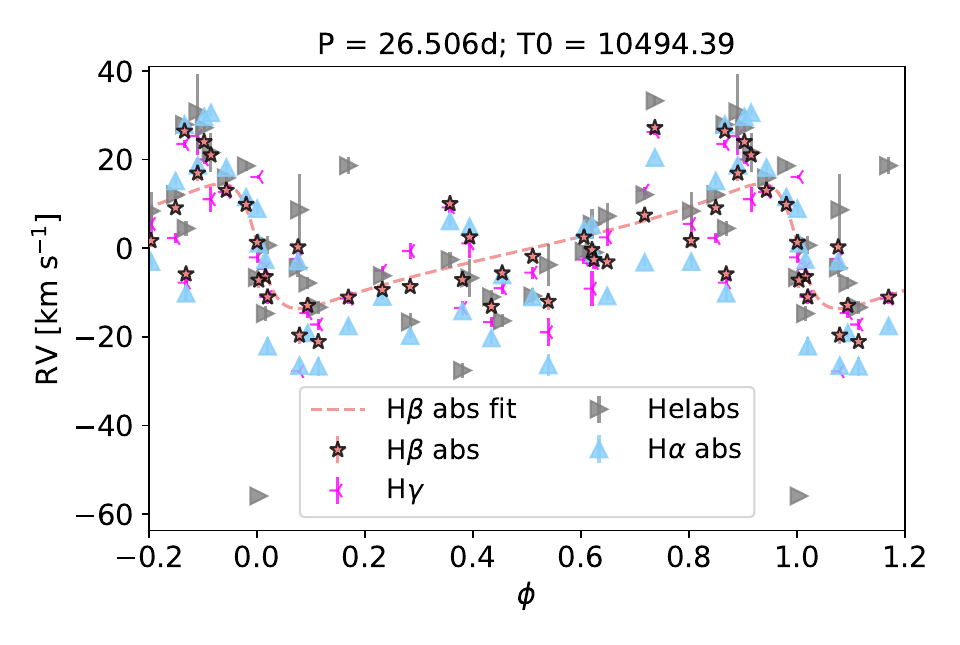}
    \end{subfigure}
     \begin{subfigure}{0.49\linewidth}
    \includegraphics[trim = 5 15 10 15, clip, width = \textwidth]{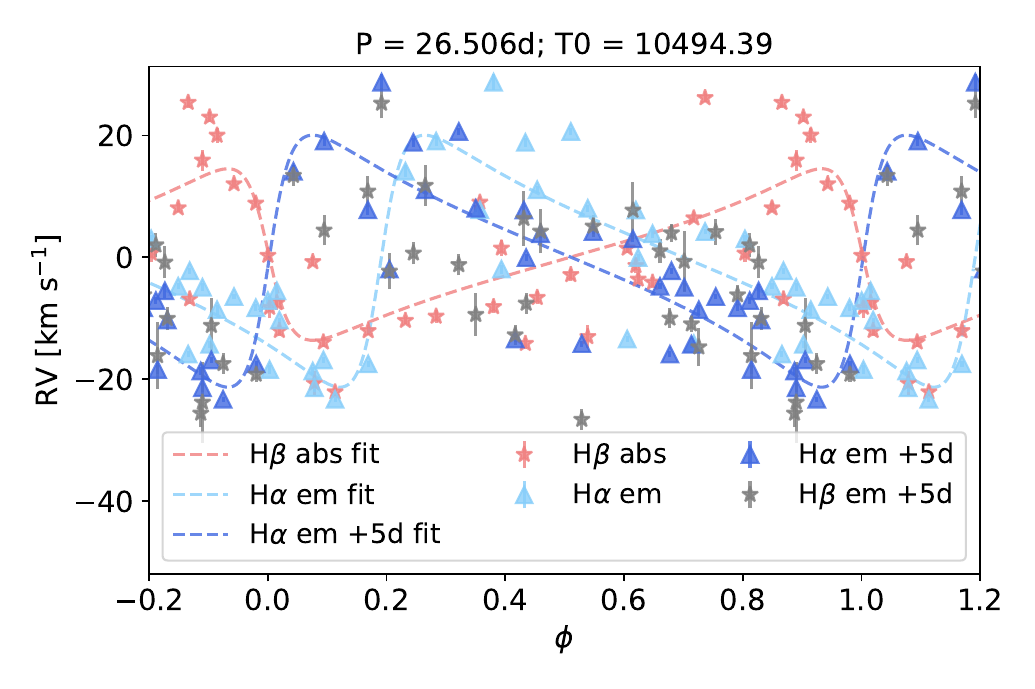}
    \end{subfigure}
    \caption{Phase-folded RVs and fits of \object{V615 Cas} with an orbital period of 26.506\,d and T0 = 10494.39. Left panel: different spectral lines, excluding H$\alpha$ emission, and the derived H$\beta$ RV curve. Right panel: H$\beta$ and the H$\alpha$ emission component with both RV curves. Darker-blue triangles show H$\alpha$ with a 5\,d difference in T0 (as listed in Table \ref{table_orbital_V615_Cas}). Grey stars show the RVs obtained from H$\beta$ emission, also with a 5\,d shift in T0.}
    \label{fig_RVs_V615Cas}
\end{figure*}

\subsection{Group A: strong absorption lines}\label{sec_groupA}
In this group showing prominent absorption lines, we have one system: V615 Cas, with a reported orbital period of $\sim26.5$\,d \citep[e.g.][]{Hutchings-Crampton_1981,Taylor-Gregory_1982,Dai_et_al_2016,Lopez-Miralles_et_al_2023}. The absorption component in the emission line of H$\alpha$ (see Fig. \ref{fig_spectral_V615Cas}) is strong enough such that RVs of the absorption and emission component can be measured separately.\\
The data for \object{V615 Cas} are spread over 15 years, and, at multiple times, cover large parts of the orbit within one orbital cycle of the reported 26.5\,d period. Phase-folded RVs are shown in Fig. \ref{fig_RVs_V615Cas} for different spectral lines. Orbital solutions are listed in Table \ref{table_orbital_V615_Cas}, where the final column shows the rms values. The orbital solutions with the lowest rms is H$\beta$.\\
The orbital periods and eccentricities mostly agree within 1$\sigma$, except for He {\sc i} abs, for which a slightly lower orbital period is found with a zero eccentricity. When fixing $P$, T0, and $e$ to those found for H$\beta$, the rms and reduced $\chi^2$ value have slightly increased compared to the values when not fixing $P$, T0, and $e$. However, the RVs of He {\sc i} abs do follow the same trend as those of other spectral lines (left panel of Fig. \ref{fig_RVs_V615Cas}). The origin of the preferred zero eccentricity is unknown.\\
The last three rows in Table \ref{table_orbital_V615_Cas} list orbital solution obtained for the highest-amplitude periods in the periodograms (Fig. \ref{fig_periodograms}, second row, left panel) of H$\beta$ and H$\alpha$ ($P = 31.055$\,d), and He {\sc i} abs ($P = 26.10$\,d) as the initial guess. The rms values have even improved compared to the orbital solutions found when using the X-ray orbital period. This highlights that, even though \object{V615 Cas} has the best orbital coverage of all BeXRBs in this study, finding the true orbital period without any beforehand knowledge remains very challenging. The 31\,d period could be a beat period due to a precessing disc.\\
Furthermore, there seems to be an anti-phase motion and an additional 5\,d shift in T0 (left panel of Fig. \ref{fig_RVs_V615Cas}) of the RVs from the H$\alpha$ emission wings (`H$\alpha$ em') with respect to other spectral lines. If this anti-phase orbital motion is true, and not an effect of the moving absorption component, like in LB-1 \citep{Abdul-Masih_et_al_2020}, it could be attributed to the NS companion, meaning the Be-star mass would be extremely low. More likely, this is a counteracting effect on the disc. Despite the large scatter, H$\beta$ em (grey stars in middle panel of Fig. \ref{fig_RVs_V615Cas}) might follow the same trend as H$\alpha$ em. However, the H$\beta$ emission component is too weak to obtain robust RVs and to fully test this hypothesis. \\
\\
Based on just one example in group A, both emission and absorption lines trace the orbital period. However, as highlighted above, effectively identifying the orbital motion without prior knowledge is challenging even in this well-observed system. Whether V615 Cas is a special case in group A in terms of anti-phase motion of emission and absorption lines or whether this will be true for most systems in this category remains to be investigated with more targets.

\subsection{Group S: stable emission lines with weaker absorption}\label{sec_groupS}
\begin{figure}
    \centering
    \includegraphics[trim = 0 15 0  15, clip, width=0.48\textwidth]{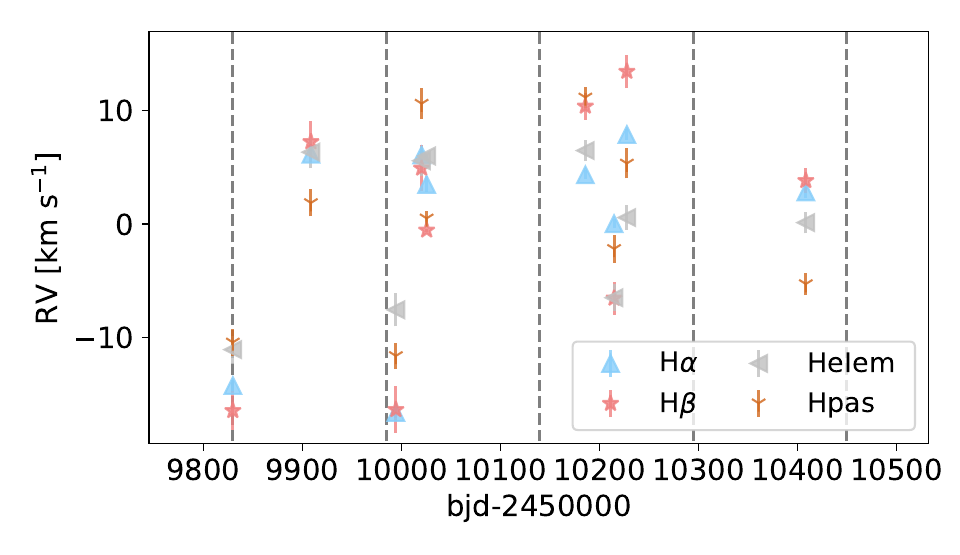}
    \caption{Radial velocities for \object{EM* VES 826}. Vertical dashed lines indicate a 155\,d period starting from the first data point.}
    \label{fig:RVs_EMVES826}
\end{figure}

\begin{figure}
    \centering
    \begin{subfigure}{\linewidth}
    \includegraphics[trim = 0 31 0 15, clip,width = \textwidth]{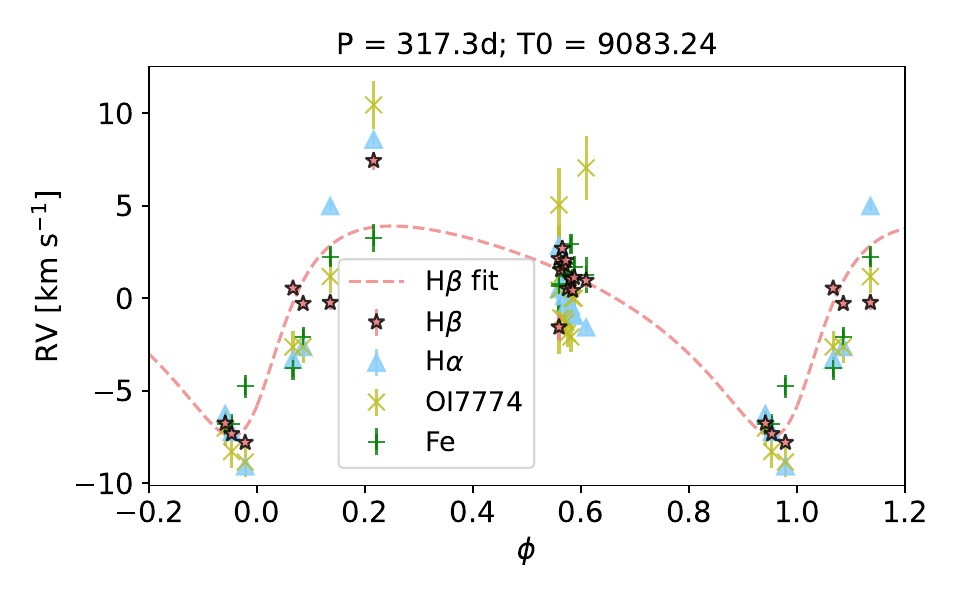}
    \end{subfigure}
     \begin{subfigure}{\linewidth}
    \includegraphics[trim = 0 15 0 10, clip,width = \textwidth]{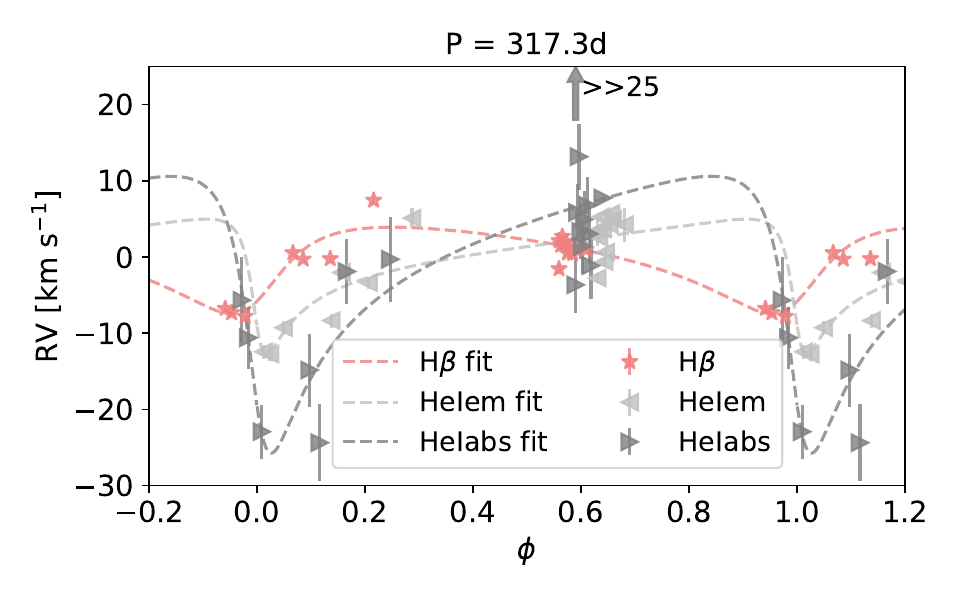}
    \end{subfigure}
    \caption{Phase-folded RVs and fits for \object{HD 259440} on a 317.3\,d period. Left panel: RVs from non-He lines and the derived H$\beta$ RV curve. Right panel: original (not fixed to the T0 of H$\beta$) RV fit for He {\sc i} em and He {\sc i} abs, with the fit of H$\beta$ as a reference. The grey arrow indicates a data point with RV\,$>$\,100\,km\,s$^{-1}$ and falls outside the range of the plot (25\,km\,s$^{-1}$).}
    \label{fig_RVs_HD259440}
\end{figure}

While the top part of the emission lines of group S systems is slightly variable, the variability is not large enough to affect the general line profile. In contrast to group A, the absorption-line signal of group S systems is rather weak. This results in a large scatter on the absorption-line RVs or the possibility that CC does not even converge (as for EM* VES 826).\\
The suggested 155\,d orbital period for EM* VES 826 is indicated by vertical dashed lines in Fig. \ref{fig:RVs_EMVES826}. The RVs of HD 259440 are shown in Fig. \ref{fig_RVs_HD259440}, phase-folded onto the 317.3\,d orbital period \citep[][]{Adams_et_al_2021}. While the orbital coverage is currently not optimal, the emission lines seem to trace the orbital period in both cases. However, more data is needed to independently determine orbital periods. Indeed, the data for EM* VES 826 are too sparse to perform orbital fitting and for HD 259440, the periastron passage is only observed during one orbital period.\\
Nevertheless, for HD\,259440 we performed orbital fitting a fixed orbital period of 317.3\,d (the X-ray orbital period). Fits for different spectral lines are listed in Table \ref{table_orbital_HD259440} and shown in Fig. \ref{fig_RVs_HD259440}.\\
Eccentricities are generally in agreement, except for He {\sc i} lines. Non-He eccentricities are below the previously reported value of 0.63 \citep{Moritani_et_al_2018} and support a lower value as also reported in \citet{Matchett_vanSoelen_2024}. The He {\sc i} abs lines are more in agreement with a value of $e = 0.63$. More data might resolve this discrepancy.\\
\\
While the signal in the absorption lines is too low to determine robust orbital solutions for systems in group S, it seems that the emission lines do trace the orbital period. Currently, the orbital coverage of both systems in this group is not sufficient to determine orbital periods independently from the X-ray orbital solutions. Nevertheless, the periodograms do show local maxima at the reported orbital periods. With more data, it is possible that these peaks become more significant.\\
Despite knowing the orbital periods, independent orbital solutions could still not be determined from these data, indicating that a large amount of data are needed to effectively determine orbital solutions for eccentric Be binaries. While this is a natural effect of eccentricity, for Be systems this issue might be more prominent because of the larger intrinsic scatter that can already be present in the RVs.


\begin{table*}
    \centering
\caption{Orbital solutions for \object{HD 259440} using different spectral lines.}
    \begin{tabular}{ ccccccccc}
     \hline
     \hline
spectral line&$P$ [d]&$e$&$\omega$ [$\degree$]&T0&$K_1$ [km\,s$^{-1}$]&$z_0$ [km\,s$^{-1}$]&rms\\
\hline
H$\beta$&317.3 $^*$&0.44\,$\pm$\,0.13&225\,$\pm$\,18&9083\,$\pm$\,11&5.68\,$\pm$\,0.75&$-$2.14\,$\pm$\,0.34&1.5\\
H$\alpha$&317.3 $^*$&0.43\,$\pm$\,0.07&278.6\,$\pm$\,7.9&9118.9\,$\pm$\,5.4&8.21\,$\pm$\,0.64&$-$0.50\,$\pm$\,0.30&0.86\\
OI&317.3 $^*$&0.29\,$\pm$\,0.17&307\,$\pm$\,42&9142\,$\pm$\,37&9.7\,$\pm$\,3.4&$-$5.1\,$\pm$\,1.1&2.6\\
He {\sc i} em&317.3 $^*$&0.74\,$\pm$\,0.64&126\,$\pm$\,107&9061\,$\pm$\,14&8.9\,$\pm$\,9.1&2.8\,$\pm$\,4.3&2.7\\
He {\sc i} abs&317.3 $^*$&0.63\,$\pm$\,0.10&131\,$\pm$\,49&9073\,$\pm$\,16&18.2\,$\pm$\,3.1&3.7\,$\pm$\,1.7&5.2\\
Fe&317.3 $^*$&0.38\,$\pm$\,0.13&240\,$\pm$\,16&9102.4\,$\pm$\,8.4&5.42\,$\pm$\,0.51&$-$0.66\,$\pm$\,0.24&0.75\\
\hline    
\end{tabular}
         \flushleft
    \begin{tablenotes}
      \small \textbf{Notes} 
      Listed uncertainties correspond to 1$\sigma$. ${^*}$ Fixed to value obtained in \citet{Adams_et_al_2021}.
    \end{tablenotes}
\label{table_orbital_HD259440}
\end{table*}

\subsection{Group LV: large-scale emission-line variability}\label{sec_groupLV}
It can be expected that large-scale variability in the emission lines affects the obtained RVs. While this should not affect pure absorption lines (i.e. absorption lines without any contamination from emission), the signal in the absorption lines of the systems in this group is often too weak. We discuss the cases of subdivisions LV-vr (extreme V/R variability; V831 Cas) and LV-ls (strong line strength and width variation; V725 Tau, X Per), and discuss V420 Aur separately. 

\begin{figure}
    \centering
    \includegraphics[trim = 0 15 0 15, clip, width=0.95\linewidth]{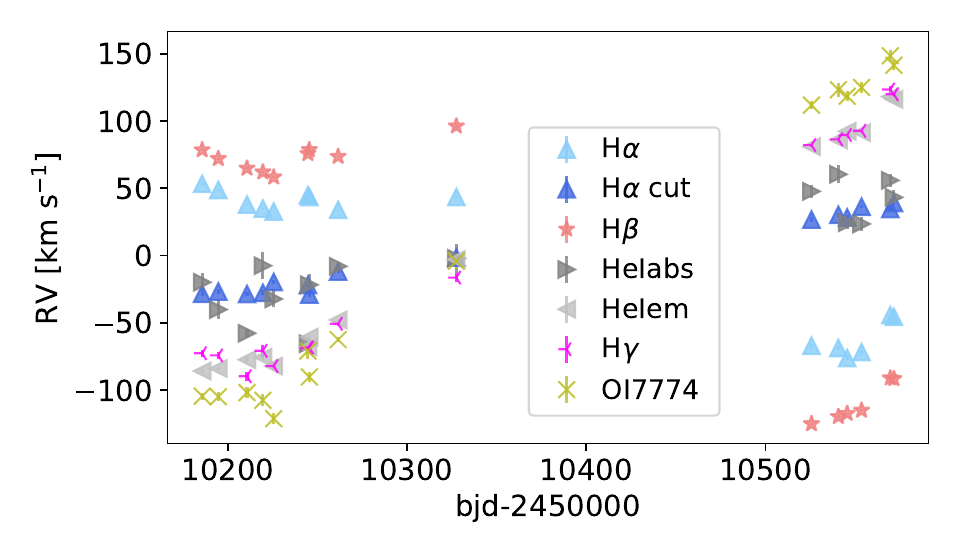}
    \caption{Non-phase-folded RVs for V831 Cas.}
    \label{fig_RVs_V831Cas}
\end{figure}

\subsubsection{Group LV-vr: extreme V/R variability}\label{sec_group-LVvr_V831Cas}
In this group we have one system: V831 Cas. The V/R variability is so strong that the emission-line profiles vary from P-Cygni-like to reverse P-Cygni-like (see Fig. \ref{fig_spectral_V831Cas}). Only in H$\alpha$ does the lower peak of the V/R variability never reach the continuum level (in our series of spectra). In such cases, it becomes challenging or almost impossible to exclude the varying top region (i.e. the emission peaks) from the fitting.\\
We can expect that, using the same template for all spectra will result in blue-shifted (red-shifted) RVs for emission-dominated (absorption-dominated) lines in spectra with V/R\,$>$\,1 compared to spectra with V/R\,$<$\,1. 
\\
This effect can indeed be seen in Fig. \ref{fig_RVs_V831Cas}. Here, even the He\,{\sc i} absorption lines apparently not contaminated by emission (marked by `a' in Table \ref{table_spectral_lines_alltargets} and `HeIabs' in Fig. \ref{fig_RVs_V831Cas}) seem to follow the red-shifted trend upon V/R reversal, even within 5$\sigma$. This suggests that they could also be affected by emission, or that the change in V/R is related to true orbital motion. Interestingly, the wings of the H$\alpha$ line (`H$\alpha$ cut'), which seem to be less variable in shape, appear to follow the trend of the absorption lines. Why the wings of H$\alpha$ would trace the motion of the absorption lines instead of the emission lines is non-trivial.\\
If the V/R variability and orbital motion are indeed related, the spiral disc density would be created by the passing of the NS, which was shown to be possible through hydrodynamic simulations \citep{Panoglou_et_al_2016, Panoglou_et_al_2018_binarydiscs}. However, it is uncertain if a 1240\,d period \citep{Reig_et_al_V831_Cas} can result in such extreme V/R variability.\\
\\ 
A better understanding of disc mechanics and companion-disc interactions \citep[as e.g. in][]{Rubio_et_al_2025} is needed to overcome these challenges and draw any conclusions on the orbital periods of systems with extreme V/R variability. Hence, currently, no orbital solution can be derived for these systems.

\begin{figure}
    \centering
    \begin{subfigure}{\linewidth}
    \includegraphics[trim = 0 31 0 15, clip, width = \textwidth]{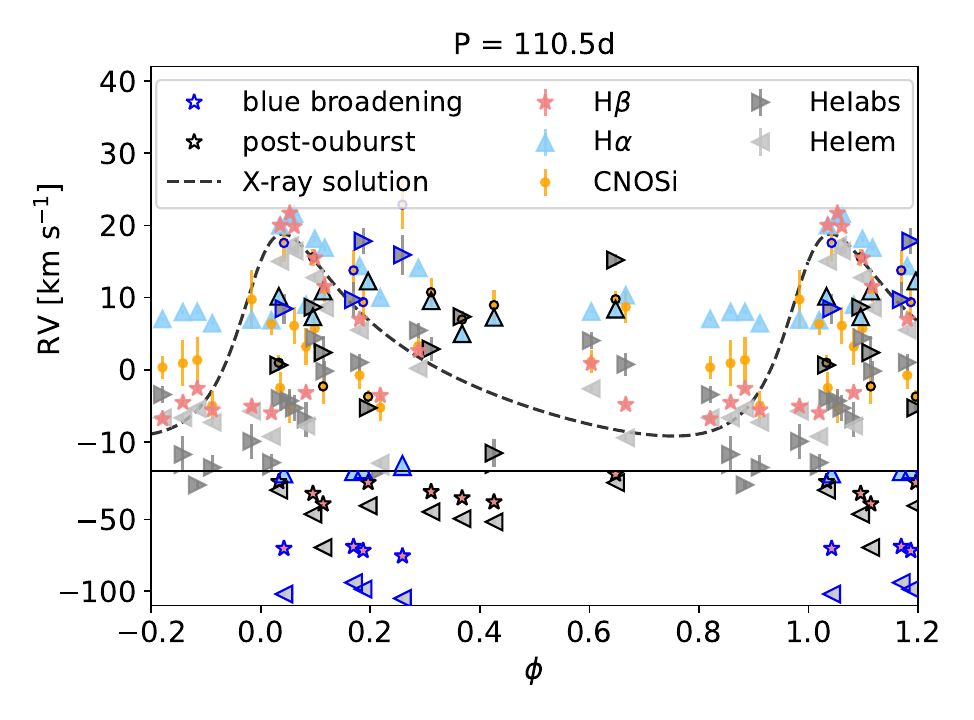}
    \end{subfigure}
    \begin{subfigure}{\linewidth}
    \includegraphics[trim = 0 10 0 10, clip, width=\textwidth]{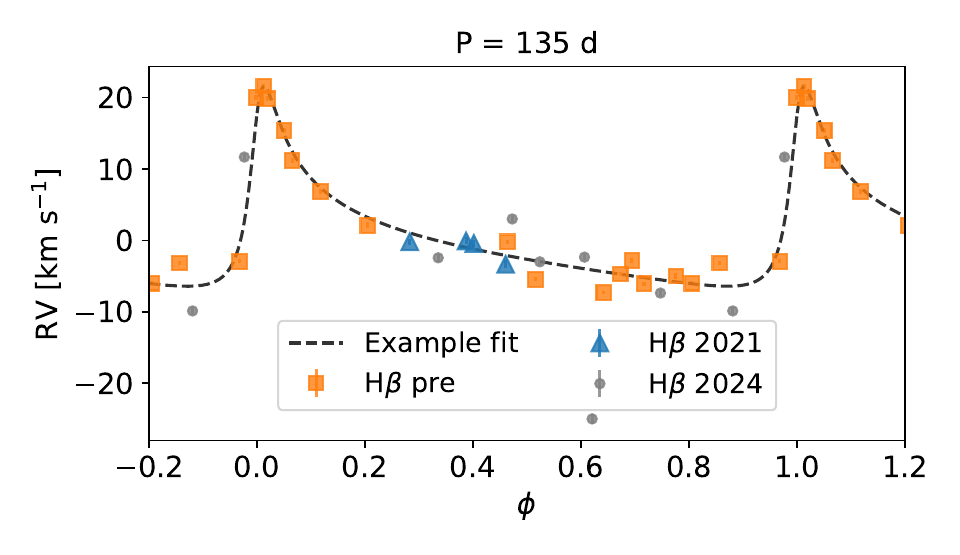}
    \end{subfigure}
    \caption{Phase-folded RVs of \object{V725 Tau}. Top: Different spectral lines and orbital solution using the X-ray period and eccentricity \citep[solid black line, $P = 110.5$\,d, $e = 0.47$;][]{Finger_et_al_1994}. Data points with blue edges suffer from blue broadening and black edges represent RVs post-blue broadening and -outburst. The lower panel is a continuity from the upper panel but is differently scaled for clarity. Bottom: Combined H$\beta$ RVs pre- and post-outburst. Orange squares are pre-outburst, blue triangles are 2021 post-outburst (blue broadened), and grey dots are 2024 post-outburst. The dashed line is a by-eye-example orbit with $P = 135$\,d and $e = 0.72$.}
    \label{fig_RV_V725_Tau_threepart}
\end{figure}
\begin{figure}
    \centering
    \begin{subfigure}{\linewidth}
        \includegraphics[trim= 0 31 0 15,clip,width=\linewidth]{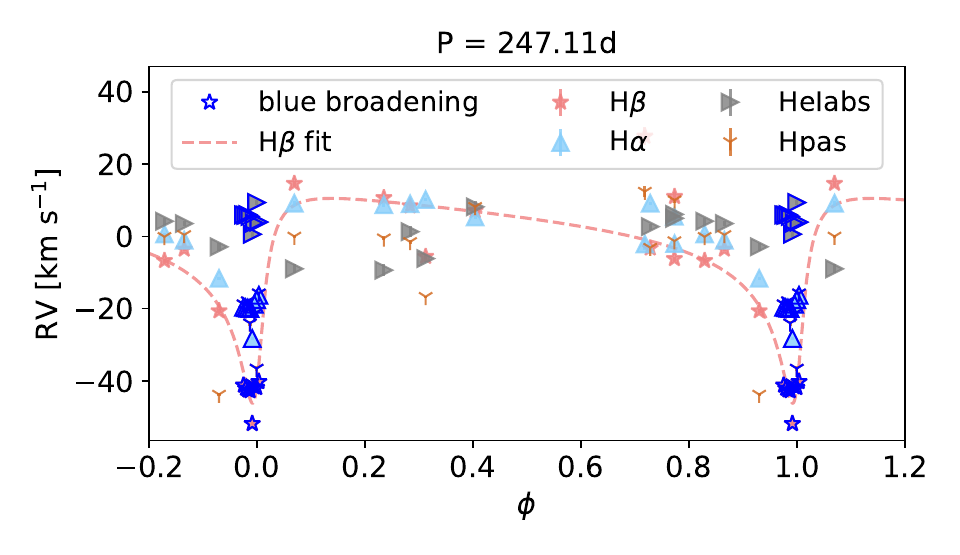}
    \end{subfigure}
    \begin{subfigure}{\linewidth}
        \includegraphics[trim= 0 15 0 15,clip,width=\linewidth]{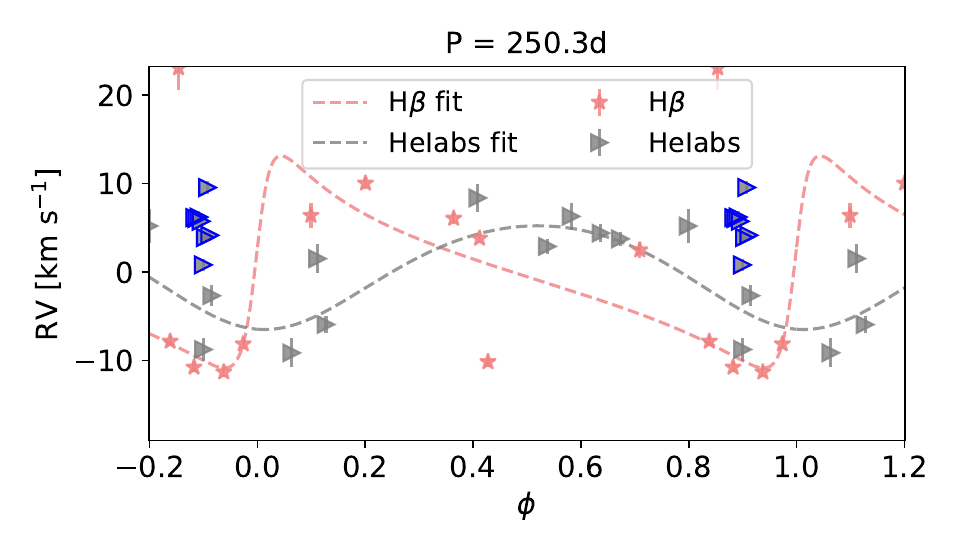}
    \end{subfigure}
    \caption{Phase-folded RVs of \object{X Per}. Blue-edged symbols correspond to blue-broadened spectra. Top: on a 247.11\,d period, with the fitted RV curve on H$\beta$.  Bottom: on a 250.3\,d period obtained when excluding the blue-broadened data points from the fit.}
    \label{fig_X_Per_RVs}
\end{figure}

\subsubsection{Group LV-ls: strong non-V/R variability}
The targets in this group (\object{V725 Tau} and \object{X Per}) show strong emission-line variation in both the width and strength of the emission lines, on top of potential V/R variability. Both systems show similar behaviour in their variability. In some spectra, there is a broadening of the emission lines, most apparent in the blue wing. This blue broadening will naturally affect the RVs determined from these spectra. While for V725 Tau the strong line-profile variability could be related to a major outburst, for \object{X Per} there is no previous outburst reported. Furthermore, \object{V725 Tau} also shows a large decrease in line strength for other spectra post-outburst.\\
In Figs. \ref{fig_RV_V725_Tau_threepart} and \ref{fig_X_Per_RVs}, the RVs measured on spectra with the apparent blue broadening (blue-edged symbols) are much more blue-shifted than the other RVs, as expected. 
Even absorption-line RVs seem to be affected. For V725 Tau, the other RVs post-outburst (black edged symbols in Fig. \ref{fig_RV_V725_Tau_threepart}) also do not follow the RVs pre-outburst. This shows the effect of large-scale variability on the obtained RVs of Be emission lines. Unfortunately, due to the low S/N in the absorption lines, we cannot derive an orbital solution from these RVs either and, for X Per, it seems like the He {\sc i} absorption lines are also affected.\\
It becomes challenging to directly combine all data to determine orbital solutions due to large, potentially untrue, RV shifts. Currently, more data is needed to understand if the apparent blue-shift of emission lines is also related to a true RV shift. However, more likely, disc mechanisms (e.g. a spiral density structure or the outburst affecting the disc, ...) are at play.\\
Therefore, to combine all data, we could split them in parts and use CC on each set of spectra. Then, each set would have a different $z_0$ and could be scaled to perform a combined fit. An example is shown in Fig. \ref{fig_RV_V725_Tau_threepart} for \object{V725 Tau}, where we split the data in pre-outburst, blue broadened spectra post-outburst of 2021, and the other set of spectra taken post-outburst in 2024. This is not a formal fit, but rather an example of a by-eye-fit as the post-outburst data does not cover enough critical parts in the orbit to perform a combined fit.\\
For \object{X Per}, the time-span of the blue-broadened data is only 10 days, so a scaling is not possible. Excluding the blue-broadened RVs from the fitting results in large uncertainties on the obtained parameters (bottom two rows in Table \ref{table_orbital_XPer}). Without fixing the orbital period, it could still be retrieved.\\
\\
In both cases, the orbital X-ray solution is not retrieved and the absorption lines do not trace the same motion as the emission lines. Thus, also for systems of group LV-ls, determining orbital periods is challenging and a vast amount of data pre- and post-changes is required.\\
Moreover, for \object{V725 Tau}, the motion traced by H$\alpha$ has a smaller semi-amplitude than the one traced by H$\beta$. This shows again that H$\alpha$ is not always a trustworthy tracer of orbital motion. Why this trend is only seen in H$\alpha$ remains unknown.

\begin{table*}
    \centering
\caption{Orbital solutions for \object{X Per} for different spectral lines.}
    \begin{tabular}{ cccccccc}
     \hline
     \hline
spectral&&&&&&&\\
spectral line&$P$ [d]&$e$&$\omega$ [$\degree$]&T0&$K_1$ [km\,s$^{-1}$]&$z_0$ [km\,s$^{-1}$]&rms\\
\hline
H$\beta$&247.11 $\pm$ 0.75&0.78 $\pm$ 0.04&216 $\pm$ 10&10138.6 $\pm$ 5.4&28.3 $\pm$ 2.2&$-$12.0 $\pm$ 1.9&8.6\\
H$\alpha$&248.13 $\pm$ 0.47&0.78 $\pm$ 0.04&214.6 $\pm$ 9.6&10145.2 $\pm$ 3.4&16.8 $\pm$ 1.2&6.6 $\pm$ 1.1&2.8\\
\hline
H$\beta$$^{\text{r}}$&250.3${^*}$&0.68 $\pm$ 0.79&98 $\pm$ 83&10359 $\pm$ 19&12 $\pm$ 23&$-$6.4 $\pm$ 5.5&4.8\\
He {\sc i} abs$^{\text{r}}$&250.3${^*}$&0.39 $\pm$ 0.38&358 $\pm$ 52&10435 $\pm$ 25&7.7 $\pm$ 4.8&$-$5.2 $\pm$ 1.4&3.2\\
\hline
\end{tabular}
         \flushleft
    \begin{tablenotes}
      \small \textbf{Notes} Listed uncertainties correspond to 1$\sigma$. $^{\text{r}}$ Orbital solution with blue broadening RVs removed for fitting. ${^*}$ Fixed to value obtained in \citet{Yatabe_et_al_2018}.
    \end{tablenotes}
\label{table_orbital_XPer}
\end{table*}

\subsubsection{V420 Aur}
\begin{figure}
     \centering
    \begin{subfigure}{\linewidth}
    \includegraphics[trim= 19 12 28 2,clip,width = \textwidth]{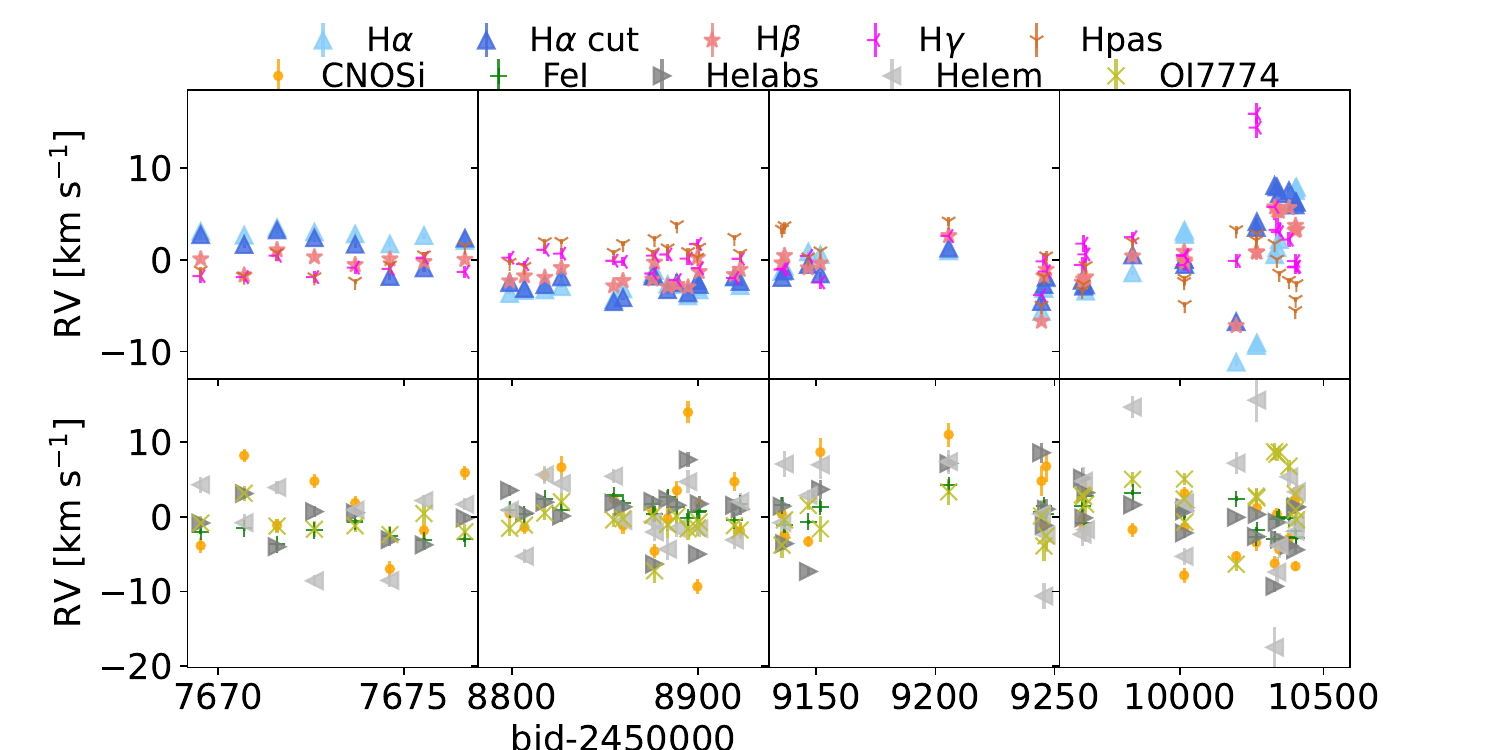}
    \end{subfigure}
    \begin{subfigure}{\linewidth}
    \includegraphics[trim= 100 15 100 195,clip,width = \textwidth]{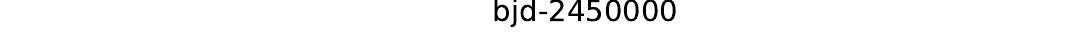}
    \end{subfigure}
    \begin{subfigure}{\linewidth}
     \includegraphics[trim = 15 15 0 15, clip, width=\linewidth]{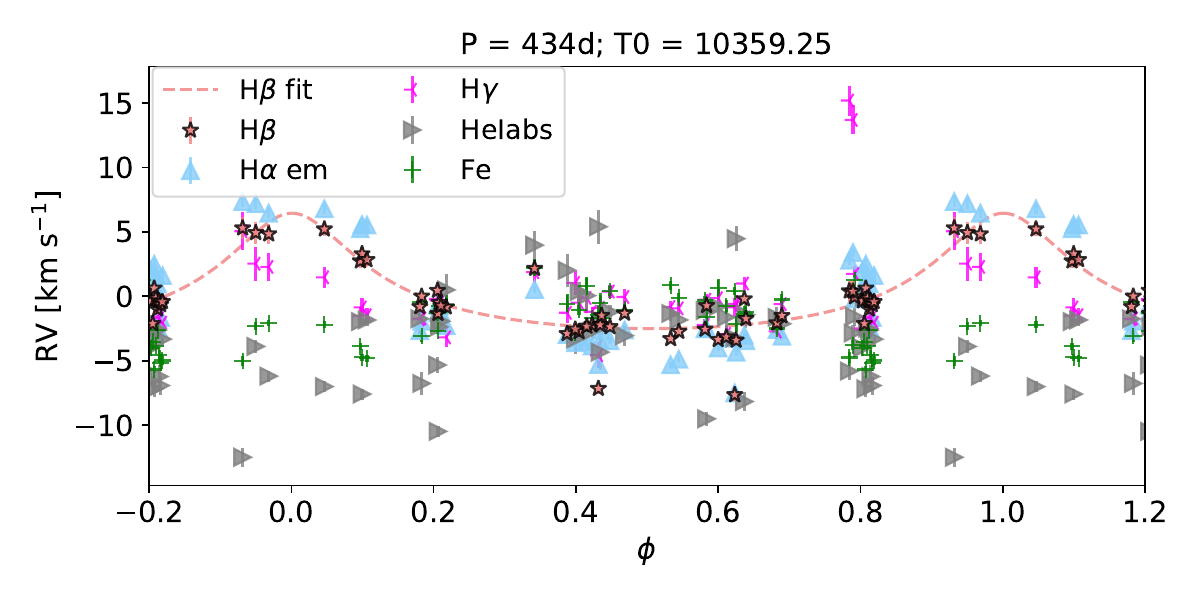}
    \end{subfigure}
    \caption{Top: Non-phase folded RVs for \object{V420 Aur} for different spectral lines, split over four panels to exclude large time gaps. Bottom: RVs phase-folded onto a 434\,d period and the fit on H$\beta$.}
    \label{fig_RVs_V420Aur}
\end{figure}

V420 Aur does not show the same variability as other systems in the LV group. Only in the final spectra does the emission-line strength seem to change (Fig. \ref{fig_spectral_V420Aur}), but without clear sign of blue broadening or large V/R variability.\\
The top panel of Fig. \ref{fig_RVs_V420Aur} shows the non-phase-folded RVs with different spectral lines. Most of the variability for the He {\sc i} lines and CNOSi blend falls within the estimated precision listed in Table \ref{table_chi2values} ($\sim$5\,km\,s$^{-1}$). Hence, apart from the most recent RV measurements in the last frame (i.e. those coinciding with the increase in line strength), not much RV variability is seen.\\
No orbital period has been reported for \object{V420 Aur}. However, the periodograms of the emission lines returned a period of $434.26\pm6.89$\,d (top-right panel in Fig. \ref{fig_periodograms}). Phase-folded RVs to this period are shown in the bottom panel of Fig. \ref{fig_RVs_V420Aur} with a tentative orbital solution on H$\beta$ resulting in $P$\,=\,433.6\,$\pm$\,3.6\,d, $e$\,=\,0.44\,$\pm$\,0.12, $\omega$\,=\,359\,$\pm$\,15\,$\degree$, T0\,=\,10359\,$\pm$\,15, $K_1$\,=\,4.5\,$\pm$\,1.1\,km\,s$^{-1}$, and rms\,=\,1.4.\\
The strong emission-line (H$\beta$, H$\alpha$, partly H$\gamma$) RVs at periastron are those corresponding to the final spectra, for which the line strength increased. Hence, the reliability of the RV change of these data points is questionable. The RVs of He\,{\sc i} and Fe (even though it is an emission line) do not follow the motion of the strong emission lines. More data at the suggested periastron passage is needed to confirm or reject the emission-line periodicity.

\section{Discussion}\label{sec_Discussion}
In Sect. \ref{subsec_Halpha_nontrend} we compare the usage of emission lines and absorption lines. We discuss implications of our derived orbital solutions on the mass function and hence the mass and nature of the companion to the Be star in Sect. \ref{sec_massfunc}. 

\subsection{Emission lines vs. absorption lines}\label{subsec_Halpha_nontrend}
\begin{figure}
    \centering
    \includegraphics[trim= 0 20 0 30,clip,width=\linewidth]{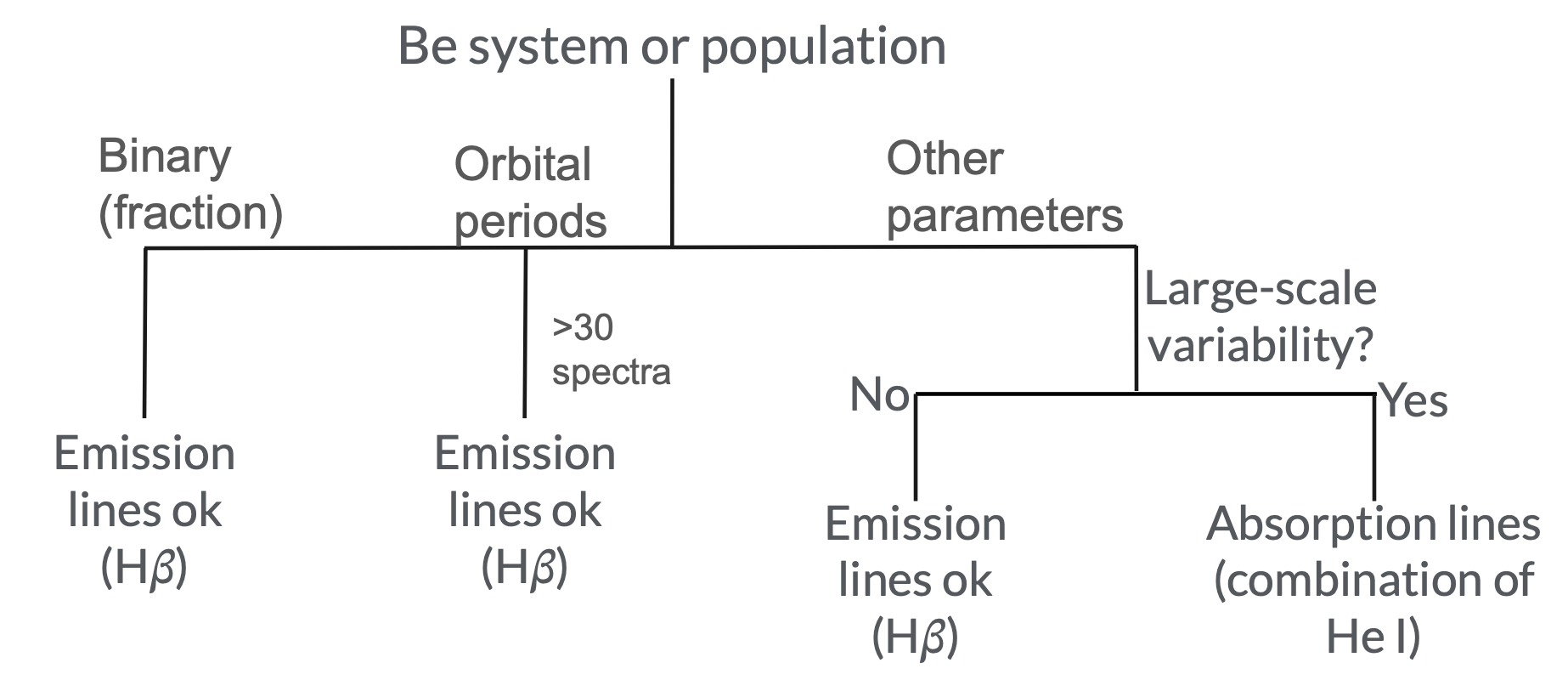}
    \caption{Schematic overview of our recommendations for the analysis of individual Be stars or populations.}
    \label{fig_schematic_suggestions}
\end{figure}

On the one hand, strong emission lines like H$\alpha$ and H$\beta$ show less short-term scatter (i.e. have a smaller $\sigma_{\text{meas}}$) compared to weaker or smeared-out lines such as the He {\sc i} absorption lines (see Sect. \ref{subsec_stability_summary}). 
On the other hand, in some systems, H$\alpha$ does not seem to trace similar motion as other lines \citep[e.g. \object{V725 Tau} discussed here or \object{LB-1};][]{Liu_et_al_2019_LB1,Abdul-Masih_et_al_2020}. The current cause is unknown, but possible scenarios could be accretion hotspots or the combined motion of an accretion disc around the companion and the Be-star disc.\\
Moreover, in some Be stars, the emission lines show large-scale variability (group LV), which drastically impacts the RV measurements. While absorption lines thus might seem like a better probe, only for two BeXRBs in our sample do absorption lines show periodicity with a larger amplitude than their uncertainties (\object{V615 Cas} and \object{HD 259440}) 
and in some other systems (e.g. \object{EM* VES 826}), CC could not even converge on the absorption lines. In some systems, it also appears that the apparent He {\sc i} absorption lines are still contaminated by disc emission (e.g. V831 Cas, Sec. \ref{sec_orb_discuss_V831Cas}). Hence, in some cases, emission lines are the only measure we have (spectroscopically).\\
Furthermore, the He {\sc i} absorption lines are more prone to spurious RV measurements (see e.g. RVs in Figs. \ref{fig_RVs_V615Cas} and \ref{fig_RVs_HD259440}). In spectra which have a low S/N, no reliable RVs can be obtained in lines with weaker and smeared out signals. Even CC does not obtain uncertainties that could indicate a spurious RV. One of the advantages of using emission lines is thus that spectra with a lower S/N can still be used to measure RVs.\\
\\
To summarise, neither absorption nor emission lines yield fully optimal results when it comes to detecting orbital motion. Below, we give an overview of our suggestions for which spectral lines to use depending on the goal (see also Fig. \ref{fig_schematic_suggestions}).\\
For the purpose of determining binary statistics in a population of Be stars or to assess if a specific system is a binary, we suggest using emission lines. Preferably, not the H$\alpha$ line, as it suffers most from variability, but instead H$\beta$ can be used. While H$\gamma$ might also be a good tracer, it is more prone to changes from emission dominated to absorption dominated when emission-line strength varies, thus requiring more caution.\\
Assuming most large-scale variability (e.g. V/R variation and blue broadening) is caused by companions, orbital periods can also be determined from the emission lines. This can be done through RV measurements, as shown here, or through the equivalent widths \citep[e.g.][]{Zamanov_et_al_2022}.\\
However, to determine accurate orbital solutions, a case-by-case study is preferred. First, one needs to determine if there is large-scale variability. If not, the signal in the emission lines will likely improve the orbital fit compared to the absorption lines. Again, we do advise against using H$\alpha$. If large-scale variability does show in the spectrum, it is best to rely more on absorption lines, as extreme V/R variability (group LS-vr; e.g. \object{V831 Cas}), or blue broadening and outbursts (group LS-lv; e.g. \object{V725 Tau}), will induce additional shifts in the RVs.

\subsection{Discussion of the orbital solutions}\label{sec_massfunc}
From the orbital solutions, we can obtain information on the mass of the unseen companion through the mass function
\begin{equation}
    f_m = \frac{m_{\text{c}}^3\sin^3 i}{(m_{\text{Be}}+m_{\text{c}})^2} = \frac{PK_{\text{Be}}^3}{2\pi G} (1-e^2)^{3/2},
\end{equation}
where the subscript `Be' and `c' refer to the Be star and companion respectively, and $K_{\text{Be}}$ is the derived $K_1$ of our orbital solutions. By using the Be-star mass and assuming an inclination of 90\,$\degree$, a minimum mass for the companion $M_{c,\text{min}}$ can be obtained. 
Below, we briefly discuss the mass function and its implications on the nature and masses of the companions for \object{HD 259440}, \object{V420 Aur}, \object{V615 Cas}. For \object{X Per}, the uncertainties are too large to obtain a meaningful mass function. We also briefly discuss the RV measurements of V831 Cas.

\subsubsection{HD 259440}
Using the solution for H$\beta$ ($P$\,=\,317.3\,d, $e$\,=\,0.44\,$\pm$\,0.13, $K_1$\,=\,5.68\,$\pm$\,0.75\,\kms), we find $f_m$\,=\,0.0044\,$\pm$\,0.0020\,$M_{\odot}$. Assuming $M_{\text{Be}}$\,=\,13.2\,(19.0)\,$M_{\odot}$ \citep{Aragona_et_al_2010}, we obtain $M_{c,\text{min}}$\,=\,0.96\,$\pm$\,0.15\,(1.21\,$\pm$\,0.19)\,$M_{\odot}$, where the errorbars will be slightly underestimated due to fixing the period. Typical NS masses are reached when 25$\degree$\,$\lesssim$\,$i$\,$\lesssim$\,45$\degree$ (30$\degree$\,$\lesssim$\,$i$\,$\lesssim$\,55$\degree$). Lower inclinations would imply a BH companion, while higher inclinations would imply a white dwarf (WD) companion. If the disc and orbit are alligned, the lack of clear absorption features in the hydrogen emission lines would indicate a moderate inclination.\\
For solutions with higher RV semi-amplitudes (e.g. H$\alpha$: $e$\,=\,0.43\,$\pm$\,0.07, $K_1$\,=\,8.21\,$\pm$\,0.64\,\kms), $f_m$ becomes 0.0134\,$\pm$\,0.0035\,$M_{\odot}$ and $M_{c,\text{min}}$\,=\,1.42\,$\pm$\,0.13\,(1.80\,$\pm$\,0.16)\,$M_{\odot}$. In this case, a WD would be ruled out.

\subsubsection{V420 Aur}
For V420 Aur, the validity of the orbital solution is based on whether we trust the RVs determined from the spectra that have stronger emission lines. If so, the orbital solution for V420 Aur ($P$\,=\,433.6\,$\pm$\,3.6\,d, $e$\,=\,0.44\,$\pm$\,0.12, $K_1$\,=\,4.5\,$\pm$\,1.1\,\kms) results in $f_m$\,=\,0.0030\,$\pm$\,0.0023\,$M_{\odot}$. Given the large uncertainty, it is difficult to make any suggestions for the nature of the companion.

\subsubsection{V615 Cas}
The H$\beta$ orbital solution ($P$\,=\,26.506\,$\pm$\,0.003\,d, $e$\,=\,0.58\,$\pm$\,0.13, $K_1$\,=\,14.1\,$\pm$\,2.6\,\kms) results in $f_m$\,=\,0.0041\,$\pm$\,0.0027. V615 Cas is classified as a B0\,V star \citep{Grundstrom_et_al_2007} and thus we can assume a mass for the Be star between 13.2-15.5\,$M_{\odot}$ \citep[e.g.][]{Janssens_mass_et_al_2022}. With $M_{\text{Be}}$\,=\,13.2\,(15.5)\,$M_{\odot}$, we obtain $M_{c,\text{min}}$\,=\,0.94\,$\pm$\,0.22\,(1.04\,$\pm$\,0.24)\,$M_{\odot}$. The detection of radio pulsations suggests the companion is a NS, and thus a much lower inclination of $i$\,$\lesssim$45-50$\degree$ is necessary. In contrast, the deep absorption or shell features observed in the spectrum indicate an almost edge-on view of the Be disc. This could imply a misalignment of the Be disc and orbital plane, as is suggested by the presence of a precession period.

\subsubsection{V831 Cas}\label{sec_orb_discuss_V831Cas}
Even though no orbital solution could be derived for V831 Cas, the absorption lines show RV variations with a semi-amplitude of $\sim50$\,\kms. If this would represent true orbital motion, the $\sim1200$\,d period \citep{Reig_et_al_V831_Cas} would lead to a minimum mass of the unseen companion $>$15\,$M_{\odot}$, even for a Be-star mass of 1\,$M_{\odot}$ (which would be far too low for its spectral type). Given the pulsar nature of the companion \citep{Haberl_et_al_1998,Reig_et_al_V831_Cas}, this orbital solution is thus not possible. Hence, the apparent absorption lines are likely contaminated by the emission component as well.

\section{Conclusion}\label{sec_Conclusion}
We performed optical spectroscopic analysis of seven BeXRBs. The goal of this work was to compare RVs obtained from different spectral lines and methods against orbital periods and potentially eccentricities obtained from the X-rays.\\ 
All three methods (CC, lpf, and bisector) can result in satisfactory results for RV determination. However, CC is more autonomous than lpf and can be used on line blends, unlike the bisector method. The measurements uncertainties of the bisector method are largely underestimated.\\
The formal precision that can be reached with strong emission lines (e.g. H$\alpha$: $\sim\,0.2-0.3$\,km\,s$^{-1}$; H$\beta$: $\sim0.5-1.5$\,km\,s$^{-1}$), is better than that of weaker (absorption) lines (e.g. He\,{\sc i}: $>3$\,km\,s$^{-1}$). Not only is there more variability in the absorption lines due to the lower signal, the pulsations of Be stars add another factor of variability. On top of that, apparent absorption lines can be contaminated by emission. Moreover, emission-line RVs are more robust for spectra with low S/N and less prone to outliers.\\
Small-scale emission-line variability in the top regions seems to cause minor issues when determining RVs. On the contrary, large-scale variability due to outbursts or V/R variation will significantly affect RV measurements. Therefore, we highlight the need for a vast amount of data. Long-term orbital campaigns can be dedicated to Be stars to monitor the orbital period in between outbursts, to circumvent extreme line-profile variability.\\
Overall, determining independent orbital periods and solutions for Be stars remains challenging. We make the following suggestions to improve the analysis of Be stars (see also Fig. \ref{fig_schematic_suggestions}):
\begin{itemize}
    \item For systems showing stable emission lines (groups A and S), we suggest that the emission lines, in particular H$\beta$, are reliable RV tracers. The signal in absorption lines of Be stars is often too weak to obtain reliable RVs. Furthermore, we advise against H$\alpha$ because its motion is known to deviate from that of other spectral lines (e.g. for \object{V725 Tau} or \object{LB-1}). The cause is currently unknown.
    \item In case of large-scale variability (group LV) due to outbursts, data pre- and post-outburst can be analysed separately and then combined. For V/R variability, this is not so trivial. If the V/R variability does not cause P-Cygni-like line-profiles, excluding the variable region (the top) can improve measurements. However, if P-Cygni like profiles occur (e.g. \object {V831 Cas}), we currently can only suggest to rely on absorption-line RV measurements.
    \item The typical criteria that are used to obtain binary fractions should be reconsidered for Be stars, as absorption lines are very prone to outliers. While, emission-lines might be a better probe for the binary statistics, caution needs to be taken about systems showing large-scale variability (group LV).
\end{itemize}
\noindent Without improvements in our understanding of how disc-companion interactions alter the spectral profiles, spectroscopic analysis on Be stars will remain challenging. Therefore, we highly encourage further studies of the mechanics of Be discs \citep[as e.g. in][]{Rubio_et_al_2025}. More specifically, we need to improve our understanding beyond solely V/R variations, but also how different spectral lines react upon interaction. Indeed, different lines can show different line-profile changes (e.g. for \object{X Per}; Fig. \ref{fig_spectral_Xper}). This is likely related to how deep in the disc the companion penetrates and the corresponding changing density profile in the disc. Furthermore, we need to understand if these interactions can cause the blue broadening seen in \object{V725 Tau} and \object{X Per} and if there is a way to correct for this. Both systems might be excellent examples to study this as they show different variability in different spectral lines. With a better understanding of companion-disc interactions, we will be able to further distinguish orbital motion from intrinsic variability.

\section*{Data availability}
The RV data are only available in electronic format at the CDS via anonymous ftp to cdsarc.u-strasbg.fr (130.79.128.5) or via \url{http://cdsweb.u-strasbg.fr/cgi-bin/qcat?J/A+A/}.

\begin{acknowledgements}
This research has used data obtained at the Mercator Observatory which receives funding from the Research Foundation – Flanders (FWO) (grant agreement I000325N and I000521N). 
This project has received funding from the FWO PhD fellowships under project 11E1721N, the Postdoctoral International Research Fellowships of Japan Society for the Promotion of Science (Graduate school of Science, Tokyo university), the European Research Council (ERC) under the European Union's Horizon 2020 research and innovation programme (grant agreements n$\degree$ 772225/MULTIPLES and 101164755/METAL), from the KU Leuven Research Council (grant METH/24/012), and from the ``La Caixa'' Foundation (ID 100010434) under the fellowship code LCF/BQ/PI23/11970035. TS acknowledges support from the Israel Science Foundation (ISF) under grant number 0603225041.

\end{acknowledgements}

\bibliography{References}

\newpage
\onecolumn
\begin{appendix}
\renewcommand{\thefigure}{A\arabic{figure}}
\setcounter{figure}{0}
\renewcommand{\thetable}{A\arabic{table}}
\setcounter{table}{0}

\raggedcolumns
\begin{multicols}{2}

\section{Short description of targets}\label{appendix_target_summary}
For each of the seven targets analysed in this work, a small overview  is given below, which focuses on previously reported (orbital) periods and eccentricities\footnote{For an overview of the high-energy aspects of the sources, the reader is referred to the listed references and references therein.}. We also describe the spectral appearance, including violet-to-red (V/R) variation, where V/R\,$>$\,1 means that the violet or blue wing dominates and V/R\,$<$\,1 indicates that the red wing is dominating. A V/R\,$\approx$\,1 corresponds to a double-peaked profile where the blue and red wing are about equally strong. An overview of the nature of important spectral lines used in this work is also listed in Table \ref{table_spectral_lines_alltargets}.

\subsection{EM* VES 826} 
\begin{figure*}
    \centering
        \includegraphics[trim= 0 15 0 40,clip,width = \textwidth]{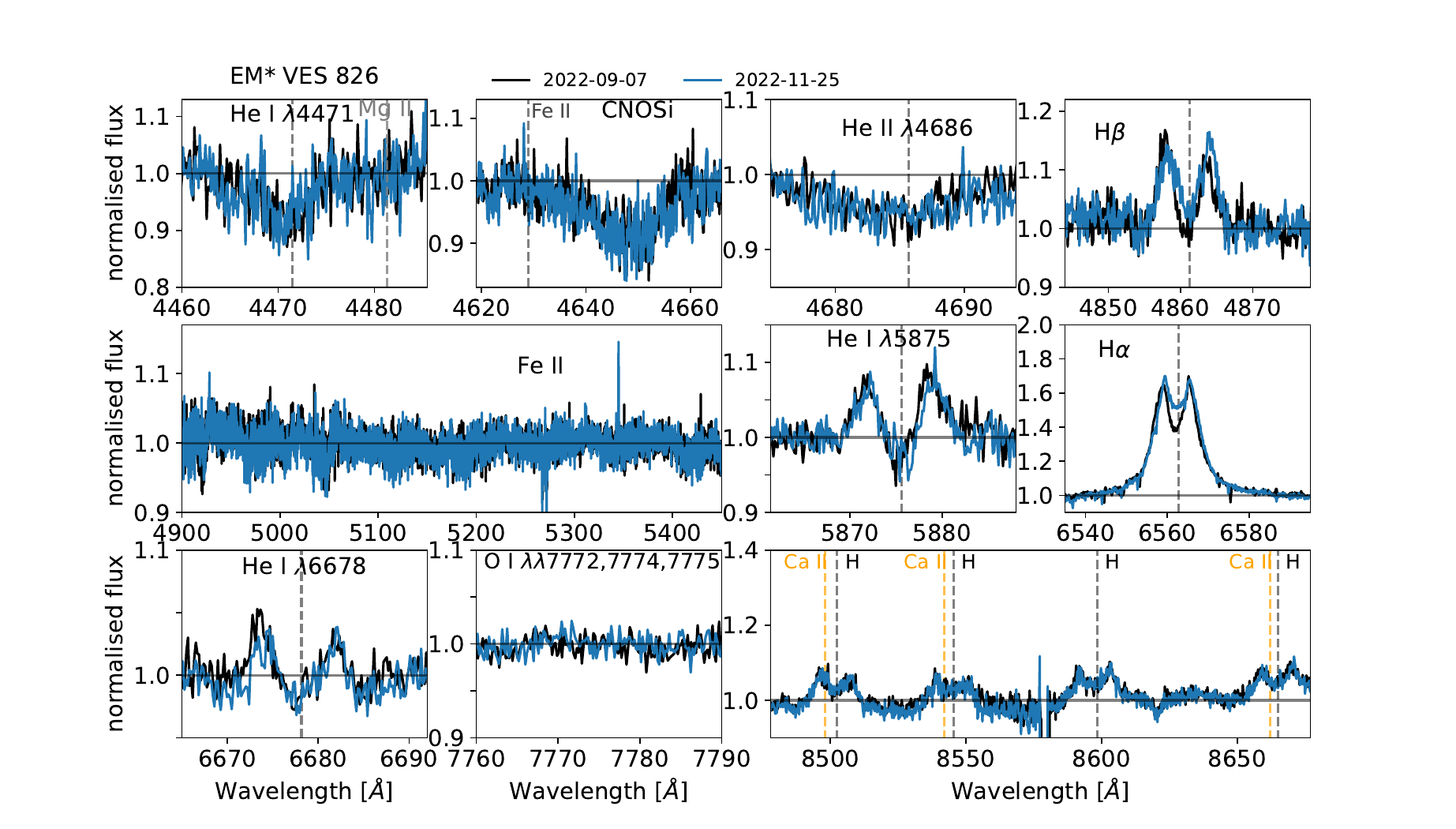}
    \caption{Spectra for \object{EM* VES 826}. The two spectra shown are the first (black) and second (blue) spectra taken, estimated to have the largest RV variability (see Fig. \ref{fig:RVs_EMVES826}).}
    \label{fig_spectral_EMVES826}
\end{figure*}

\object{EM* VES 826} (\object{LS V +44 17}, \object{RX J0440.9+4431}) contains a B0.2\,Ve star \citep{Reig_et_al_2005_VES} and is not listed in the BESS database. It was first proposed to be a Be-X-ray binary by \citet{Motch_et_al_1997} and was listed as a persistent X-ray binary with a pulse period of $P_{\text{puls}} \sim205$s \citep[][]{La_Palombara_et_al_2012, Ferrigno_et_al_2013_VES}. An orbital period of $P \sim 150-155$\,d was also estimated from the X-rays \citep[][]{Tsygankov_et_al_2012, Usui_et_al_2012_VES,Ferrigno_et_al_2013_VES}. It is one of the three BeXRBs with radio counterparts \citep{van_den_Eijnden_et_al_2024}.\\
\object{EM* VES 826} is also know to have X-ray outbursts in 2011 and 2022-2023 \citep[see][and related telegrams]{Tsygankov_et_al_VES_tele,Nakajima_et_al_2022_outburst_VES}. Spectroscopic monitoring around the 2011 outburst showed that the outburst might be connected with the size of the circumstellar disc, which was largest during the outburst and decreased right after the outburst \citep{Yan_et_al_2016_VES}. Furthermore, long spectroscopic monitoring suggest an eccentricity $\geq 0.4$ \citep{Yan_et_al_2016_VES}.\\
\\
For \object{EM* VES 826}  we acquired nine epoch spectra, taken between September 2022 and March 2024. In these spectra, \object{EM* VES 826} looks quite different than in the ones shown in \citet{Reig_et_al_2005_VES}. Both H$\alpha$ and H$\beta$ are still clear double peaked emission lines and have remained quite stable. However, H$\gamma$ also shows an emission core, which seems absent in \citet{Reig_et_al_2005_VES}. The He {\sc i} absorption lines seem weaker now compared to their spectra and some are even seen in emission. Both the CNOSi line blend and He {\sc ii} lines are seen in absorption, while some He {\sc i} lines are in emission. Fe {\sc ii} emission lines seem to be absent. A typical spectrum is shown in Fig. \ref{fig_spectral_EMVES826}.\\
Even though the spectral resolution of the data in \citet{Reig_et_al_2005_VES} is lower (5000-10000) than the Mercator data, the differences in spectral appearance are most likely related to the intrinsic variability of the disc and timing (pre- or post-outburst) of the observations.

\subsection{HD 259440}
\begin{figure*}
    \centering
        \includegraphics[trim= 0 15 0 30,clip,width = \textwidth]{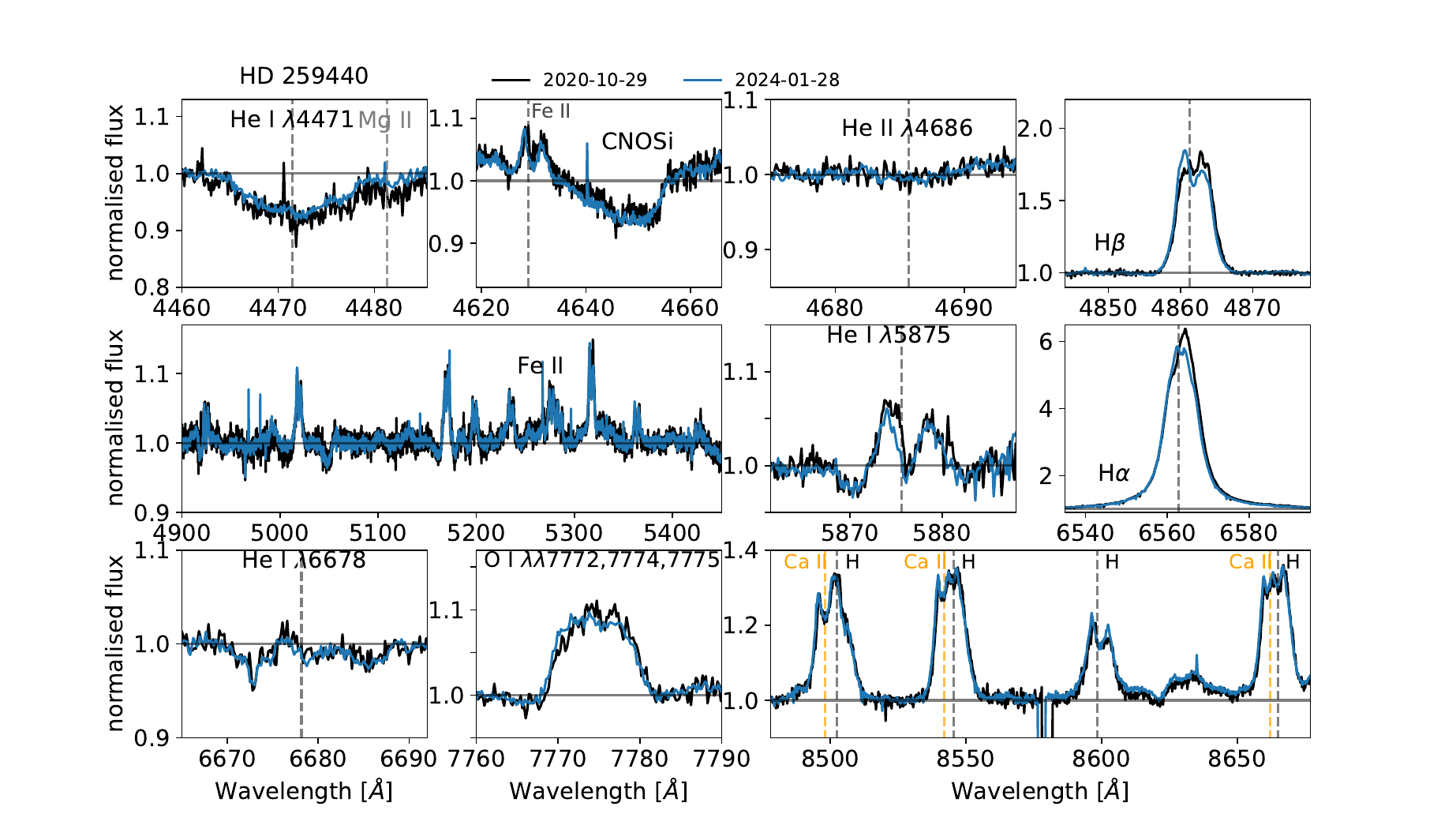}
    \caption{Spectra for \object{HD 259440}. The two spectra shown represent one spectrum close to periastron (blue) and one which is not (black, see Sect. \ref{sec_groupS}).}
    \label{fig_spectral_HD259440}
\end{figure*}

\object{HD 259440} (\object{EM* MWC 148}, \object{LS VI +05 11}) hosts a B0pe star \citep{HD259440_spt}, which has an estimated mass between 13.2\,-\,19.0$M_{\odot}$ \citep{Aragona_et_al_2010}. It shows periodic $\gamma$-ray and X-ray outbursts. No pulse period has been reported in the literature. The binary nature was first suggested by \citet{Falcone_et_al_2010} based on the X-ray spectrum. Several studies have later reported a period of $\sim$317\,d derived from the $\gamma$-ray and X-ray data \citep{Bongiorno_et_al_2011, Moritani_et_al_2018, Adams_et_al_2021}. Based on RV measurements of the H$\alpha$ emission line, a similar period was derived with a high eccentricity of $\sim$0.6 \citep{Moritani_et_al_2018}, much smaller than the previously reported $\sim$0.83 \citep{Casares_et_al_2012}. Based on the mass function, \citet{Moritani_et_al_2018} suggested that the companion is a NS with $M\le$2.5$M_{\odot}$.\\
\\
For \object{HD 259440}, the first 26 spectra were taken in November 2009 for eight consecutive days. Each night, two to three consecutive exposures of 1800s were taken. An additional 12 spectra were obtained between October 2020 and March 2024.\\
The overall spectral shape is not extremely variable over the observation time base. The H$\alpha$ line is weakly double peaked, but H$\beta$ and other emission lines clearly are double peaked. While H$\alpha$ shows variability in the top, sometimes even becoming single peaked, H$\beta$ shows much less variability. He {\sc ii} lines seem to be absent and most He {\sc i} lines are in absorption. Fe {\sc ii} emission lines are present throughout the whole spectrum, especially within the region of 5150-5380$\AA$. Interestingly, \object{HD 259440} shows stronger emission in Ca {\sc ii} $\lambda\lambda$8498,8542,8662 than the H Paschen emission lines. This line is also seen in other Be systems, where the companion is a hot subdwarf (e.g. $o$ Pup \citeauthor{Koubsky_et_al_2012} \citeyear{Koubsky_et_al_2012} and HD 55606 \citeauthor{Chojnowski_et_al_2018} \citeyear{Chojnowski_et_al_2018}). The typical spectrum is shown in Fig. \ref{fig_spectral_HD259440}.

\subsection{V420 Aur}
\begin{figure*}
    \centering
        \includegraphics[trim= 0 15 0 40,clip,width = \textwidth]{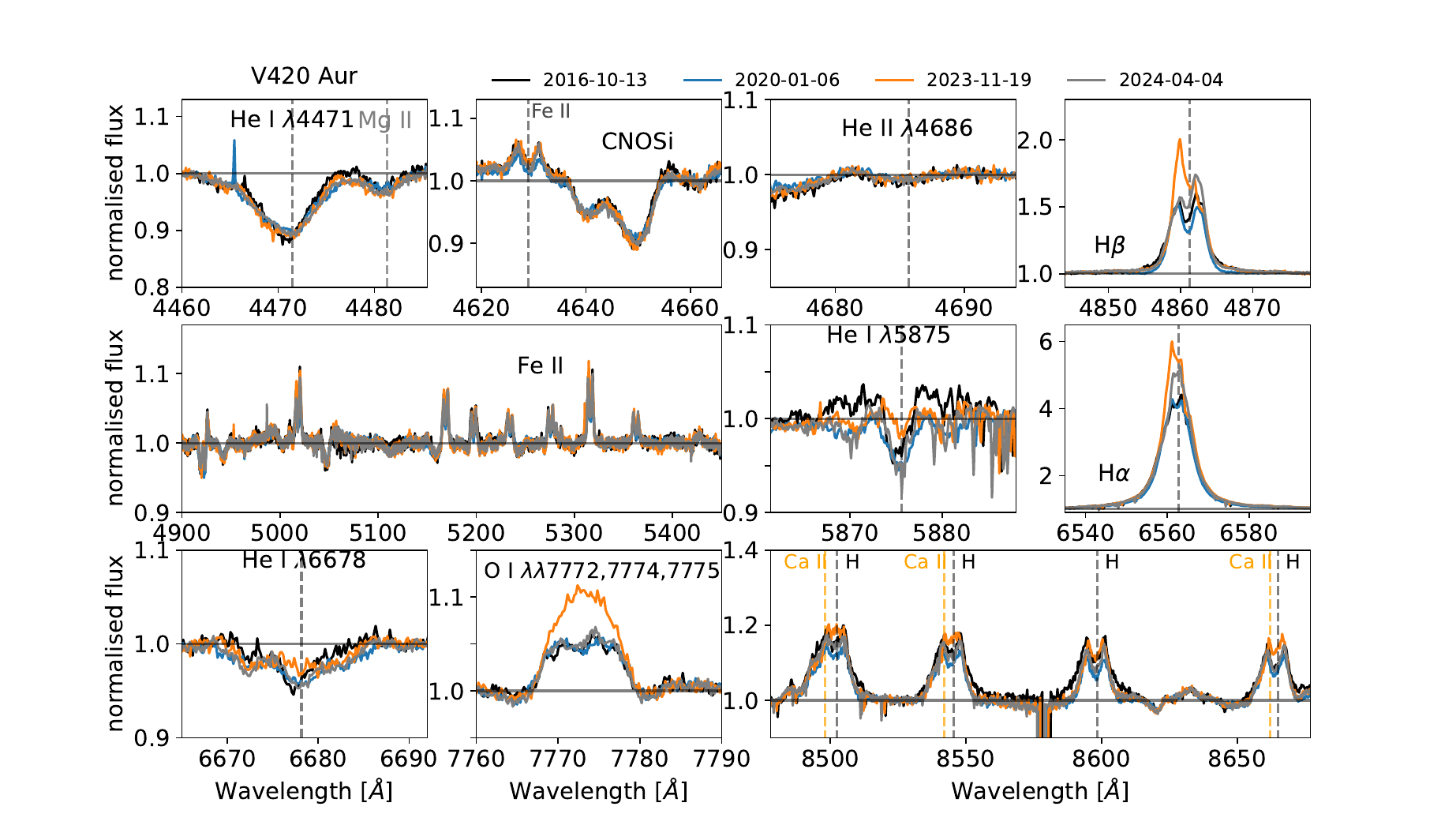}
    \caption{Three spectra of \object{V420 Aur}, taken in different years (see legend).}
    \label{fig_spectral_V420Aur}
\end{figure*}

\object{V420 Aur} (\object{HD 34921}) was first suggested to be the optical counterpart of the X-ray source \object{4U 0515+38} by \citet{Polcaro_et_al_1990}. A year later, \citet{Rossi_et_al_1991} reported on short (300s) time-scale variability in the equivalent width of the H$\alpha$ emission line and \citet{Percy_et_al_2004} report a 0.8\,d variability period based in \textit{Hipparcos} photometry. The spectral type of the Be optical component in the system is B0\,IV\,pe \citep{Polcaro_et_al_1990}. Other than that, not much information can be found in the literature about periodic variability or orbital periods.\\
\\
For \object{V420 Aur}, the first 142 spectra were taken in October 2016 for eight consecutive nights with 10-15 consecutive spectra per night with 120s exposure times. The remaining 45 epochs were taken between November 2019 and March 2024.\\
The H$\alpha$, H$\beta$, and H$\gamma$ emission lines are double peaked. Almost all He {\sc i} lines are in absorption. Some are entangled with Fe {\sc ii} emission lines, which are present throughout the whole spectrum, especially within the region of 5150-5380$\AA$. The O {\sc i} $\lambda\lambda$7772,7774,7775 line is in emission and CNOSi in absorption. An Fe {\sc ii} emission line is also seen in the bluest part of this line blend. No He {\sc ii} lines are seen. There is also emission in Ca {\sc ii} $\lambda\lambda$8498,8542,8662. However, in contrast to \object{HD 259440}, it is not stronger than the H Paschen lines.\\
The spectra appear variable in the blocks they have been taken. There are four time blocks in which spectra are taken: 2016, 2020, 2023, and 2024. There seems to be V/R variability on time scales of years and also the emission-line strength or width varies per block of observations. Between 2016 and 2020, the emission line width has decreased. Then, in 2023 and 2024, the emission line strength increased. However, the increase is not seen in every spectral line. For example, H$\alpha$ is clearly affected, but O {\sc i} $\lambda\lambda$7772,7774,7775 seems much less affected (apart from in 2023). This can all be seen in Fig. \ref{fig_spectral_V420Aur}.

\begin{figure*}
    \centering
        \includegraphics[trim= 0 15 0 40,clip,width = \textwidth]{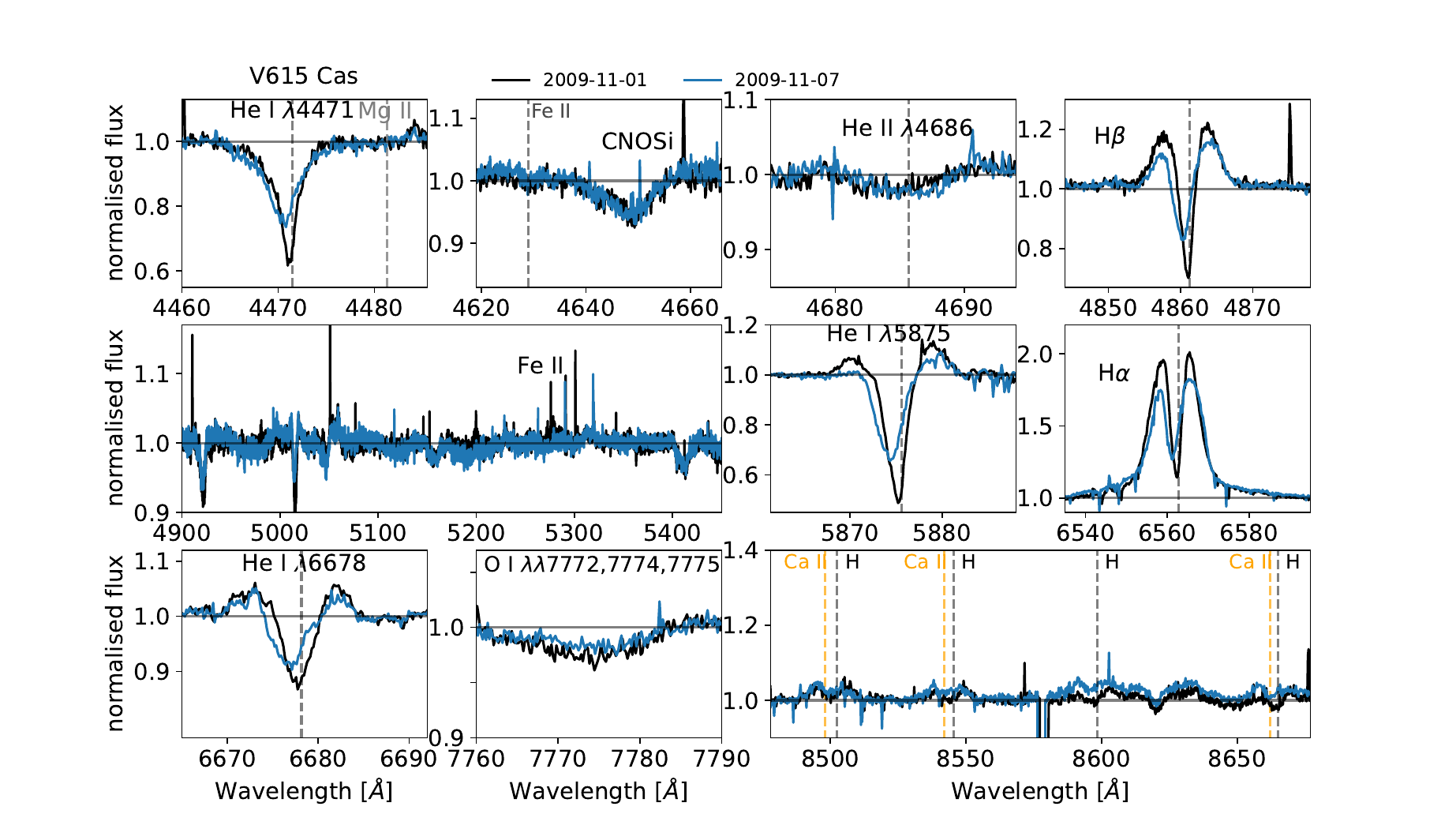}
    \caption{Spectra of \object{V615 Cas} at RV extremes. The black spectrum is close to periastron passage.}
    \label{fig_spectral_V615Cas}
\end{figure*}

\subsection{V615 Cas}
\object{V615 Cas} (\object{LS I +61 303}) hosts a B0\,V star \citep{Grundstrom_et_al_2007}. It is detected as an X-ray, radio, and $\gamma$-ray source. It has been suggested to be a microquasar with a BH of $\sim4M_{\odot}$ \citep{Punsly_1999}, which could be supported by the X-ray data \citep{Massi_et_al_2017}. However, a recent study discovered radio pulsations with a period $P_{\text{spin}} =269$\,ms \citep{Weng_et_al_2022}, indicating a spinning NS in the system. This was already suggested by multiple studies \citep[e.g.][]{Dhawan_et_al_2006,Papitto_et_al_2012}.\\
The orbital period of $\sim$26.5\,d is confirmed by several independent studies in different high-energy bands \citep[e.g.][]{Hutchings-Crampton_1981,Taylor-Gregory_1982,Dai_et_al_2016,Lopez-Miralles_et_al_2023} and also through the equivalent width of the H$\alpha$ line \citep{Zamanov_et_al_2013}. An eccentricity of $0.537 \pm 0.034$ was reported by \citet{Aragona_et_al_2009}, in line with the one reported by \citet{Casares_et_al_2005} ($e = 0.63\pm0.11$). However, based on polarimetry of the disc, the eccentricity is suggested to be much smaller \citep[$e<0.15$;][]{Kravtsov_et_al_2020}. Furthermore, the system is known to have a precession period of $\sim26.9$\,d, discovered in different data \citep[e.g.][]{Massi-Jaron_2013,Dai_et_al_2016,Wu_et_al_2018}. 
The proximity of the precession period to the orbital period leads to apparent periods or beat periods of 26.7\,d and 1667\,d \citep[e.g.][]{Massi-Jaron_2013,Massi-Torricelli_2016}.\\
\\
For \object{V615 Cas}, the first 21 spectra were taken in November 2009, with three consecutive 1800s exposures per night, thus covering 7 nights. Furthermore, 25 epoch spectra were taken between August 2022 and March 2024, where the final 16 were taken two to three nights apart over a time span of 60 days (about two orbital periods), one more spectrum was obtained in November 2024.\\
All emission lines present in the spectra of \object{V615 Cas} are double peaked and their absorption core is very strong. There is some V/R variability seen over a time span of 15 years. The absorption cores show clear motion (see Fig. \ref{fig_spectral_V615Cas}).\\
Emission lines belonging to Fe {\sc ii} or the H Paschen series are only weakly present. Most He {\sc i} lines are in absorption, as is the O {\sc i} $\lambda\lambda$7772,7774,7775 line. He {\sc ii} lines are also present in absorption. The CNOSi blend appears rather weak and is dominated by noise in some of the spectra.\\
Interestingly, the absorption lines and the absorption cores of the emission lines seem to become stronger near periastron (blue spectrum in Fig. \ref{fig_spectral_V615Cas}). In a double luminous-star binary, this would be an effect of light of one star being blocked as they move in front of one another. However, for \object{V615 Cas}, the only two light sources are the Be star and the disc. Hence, the increase in absorption-line strength could be an effect of the disc being truncated at periastron as the NS companion moves through the disc.

\begin{figure*}
    \centering
        \includegraphics[trim= 0 15 0 40,clip,width = \textwidth]{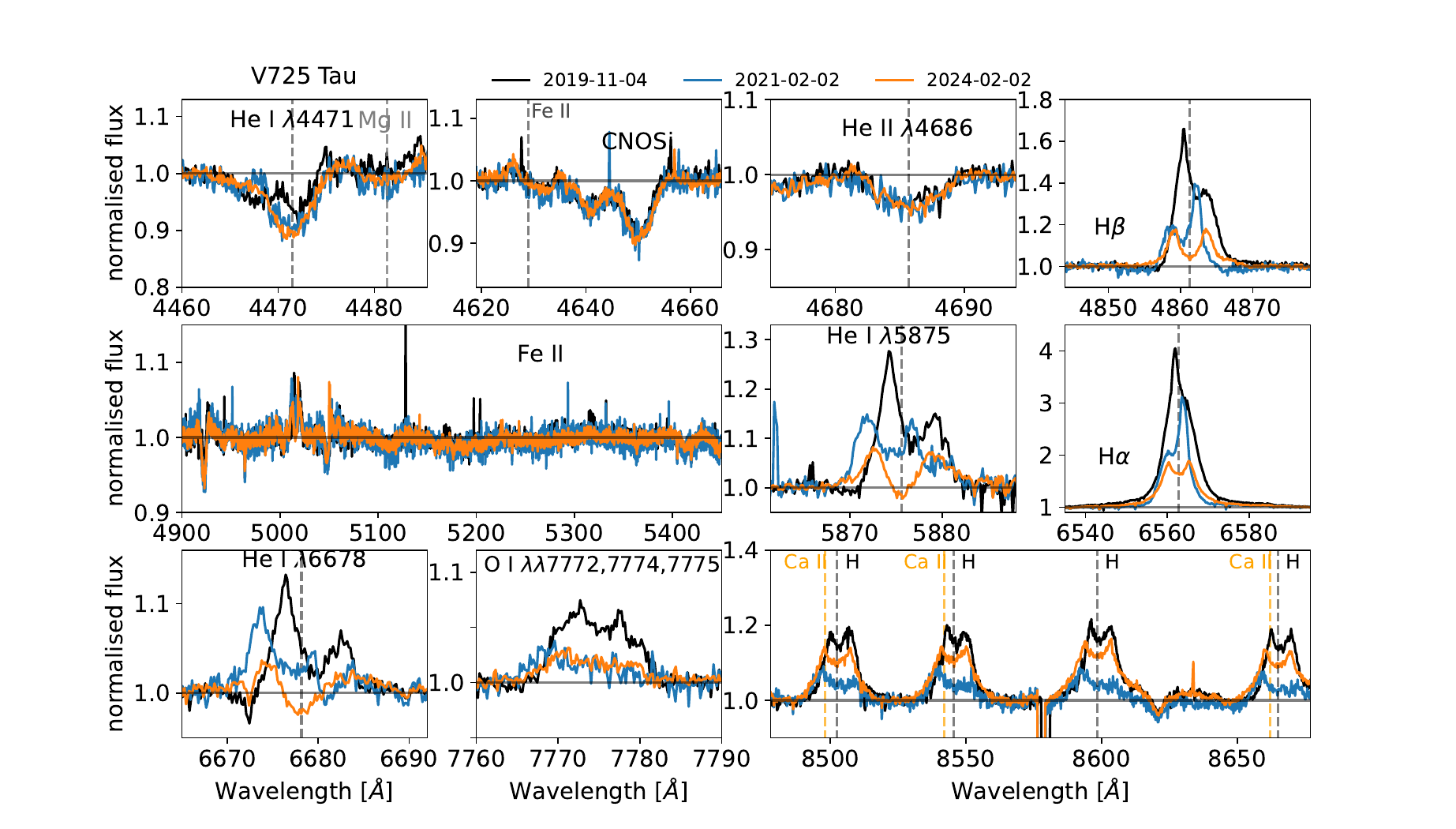}
    \caption{Spectra for \object{V725 Tau}. The black spectrum was taken before the 2020 outburst, the blue one close after the outburst, and the orange one years after the outburst.}
    \label{fig_spectral_V725Tau}
\end{figure*}

\subsection{V725 Tau}
\object{V725 Tau} (\object{HD 245770}, \object{A 0535+26}) consists of a O9.5\,III star \citep{Giangrande_et_al_1980} and a 104s X-ray pulsar \citep{Rosenberg_et_al_1975}. Several independent groups \citep{Priedhorsky-Terrell_1983,Finger_et_al_1994,Coe_et_al_2006} have derived a period of 110-111\,d based on X-ray outburst data, while \citet{Motch_et_al_1991} also confirmed this period with equivalent width measurements of H$\beta$. \citet{Finger_et_al_1994} derived an eccentricity of 0.47. By fixing the orbital period, \citet{Hutchings_1984} showed that spectroscopic data agrees with the 111\,d period, while reporting an eccentricity of $\sim$0.3.\\
Photometric analysis did not confirm the 111\,d period. However, \citet{Clark_et_al_1999} showed clear evidence for a a 103\,d and 1400\,d period in the optical light curve. \citet{Larionov_et_al_2001} attributed these periodicities to a disc precessing in a 1400\,d period, which, combined with the 111\,d orbital period, would result in the 103\,d beat period seen in the light curves. It is believed that the precessing disc is the tilted Be decretion disc and does not support evidence for a NS accretion disc \citep{Haigh_et_al_2004}. Furthermore, a 500\,d periodicity is detected in the V/R variations of \object{V725 Tau} \citep{Moritani_et_al_2010_Tau}, which is attributed to a one-armed density wave in the disc.\\
\object{V725 Tau} underwent a major X-ray and radio outburst in November 2020 (see e.g. \href{https://www.astronomerstelegram.org/?read=14193}{this telegram}), which is after the first 17 spectra were taken. All other spectra are taken after January 2021 and thus after the major outburst. Furthermore, \object{V725 Tau} is reported to have had several major X-ray outbursts starting from 1994 \citep[see \href{https://www.astronomerstelegram.org/?read=14193}{related telegrams} and][]{Finger_et_al_1996}.\\
\\
For \object{V725 Tau}, 22 spectra are available. Of those, 17 spectra were taken before the 2020 outburst in November 2019 to March 2020, four were taken a few months after the outburst (January and February 2021), and the other eight were taken between March 2022 and April 2024. \\
The spectra of \object{V725 Tau} are very variable over the 5\,yr time period of the observations (see Fig. \ref{fig_spectral_V725Tau}). In the most recent spectra of 2024, the emission lines are much weaker than in earlier ones, which causes, for example, the emission in H$\gamma$ to dominate the line in the earliest spectra and only be present in the core of the absorption line in the ones of 2024. There is also clear V/R variation. Moreover, in the spectra of 2021, all emission lines either seem to have blueshifted or part of the red wing is no longer visible (which we deem more likely). This phenomenon also coincides with those spectra that have V/R $<$ 1 (i.e. a stronger red emission, than blue) and the first ones taken right after the 2020 outburst (January and February 2021), while the spectra of 2019 have V/R $>$ 1 and those of 2022 and later show V/R $\approx$ 1. Unfortunately, our spectra do not cover the period between March 2021 and March 2022, such that the change in V/R seems rather abrupt in our data set. However, this is not necessarily the true case.\\
The absorption lines of He {\sc i} and CNOSi are clearly present and He {\sc ii} absorption is also observed. Fe {\sc ii} emission lines are rather weakly present. 

\begin{figure*}
    \centering
        \includegraphics[trim= 0 15 0 40,clip,width = \textwidth]{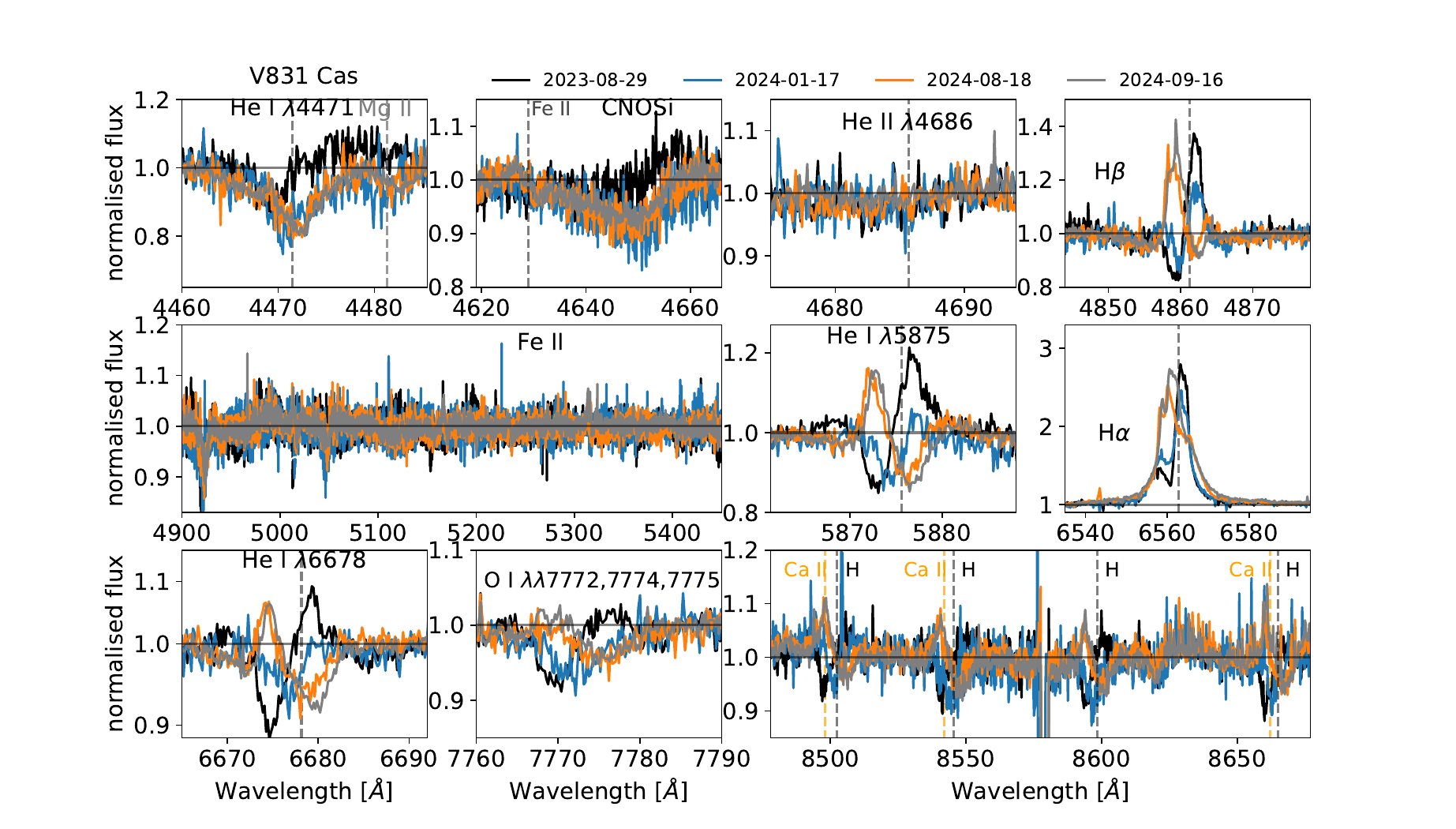}
    \caption{Spectra of V831 Cas showing clear V/R variability over the course of one year.}
    \label{fig_spectral_V831Cas}
\end{figure*}

\subsection{V831 Cas}\label{sec_appendix_V831Cas}
The Be star in \object{V831 Cas} (\object{LS I +61 235}, \object{RX J0146.9+6121}) has a B1\,IIIe spectral type \citep{Motch_et_al_1997}. It is known to be the slowest pulsar, with a pulse period $\sim$1400\,s \citep{Haberl_et_al_1998,Reig_et_al_V831_Cas}.\\
\citet{Sarty_et_al_2009} found three photometric periods less than a day, which they attributed to stellar oscillations from the Be star and its spin period. Furthermore, infrared and optical spectroscopy revealed the presence of a one-armed density wave in the Be disc with a period of 1240\,d \citep{Reig_et_al_V831_Cas}, which causes strong V/R variations. No orbital period has been reported yet.\\
\\
For \object{V831 Cas}, the 15 epoch spectra were obtained between August 2023 and September 2024. These spectra show clear V/R variability over the course of one year (see Fig. \ref{fig_spectral_V831Cas}). The extend of the variability is large enough to make all emission lines (including He {\sc i} and O {\sc i} $\lambda\lambda$7772,7774,7775), except H$\alpha$, change shapes from an apparent P-cygni profile to an apparent reverse P-cygni profile (i.e. a mirror flip of the spectral lines) over the one-year baseline of the spectra. Almost all He {\sc i} lines show emission. There are Fe {\sc ii} emission lines are weakly present and He {\sc ii} absorption lines seem absent.\\
In the most recent spectra (August and September 2024, orange and grey), the H$\alpha$ and H$\beta$ profiles are distorted, with not only a V/R $>>$ 1, but also a triple-peaked emission profile. According to \citet{Stefl_et_al_2007}, this occurs in the transition period from V/R $>$ 1 to V/R $<$ 1, which could indeed be the case here.

\begin{figure*}
    \centering
        \includegraphics[trim= 0 15 0 40,clip,width = \textwidth]{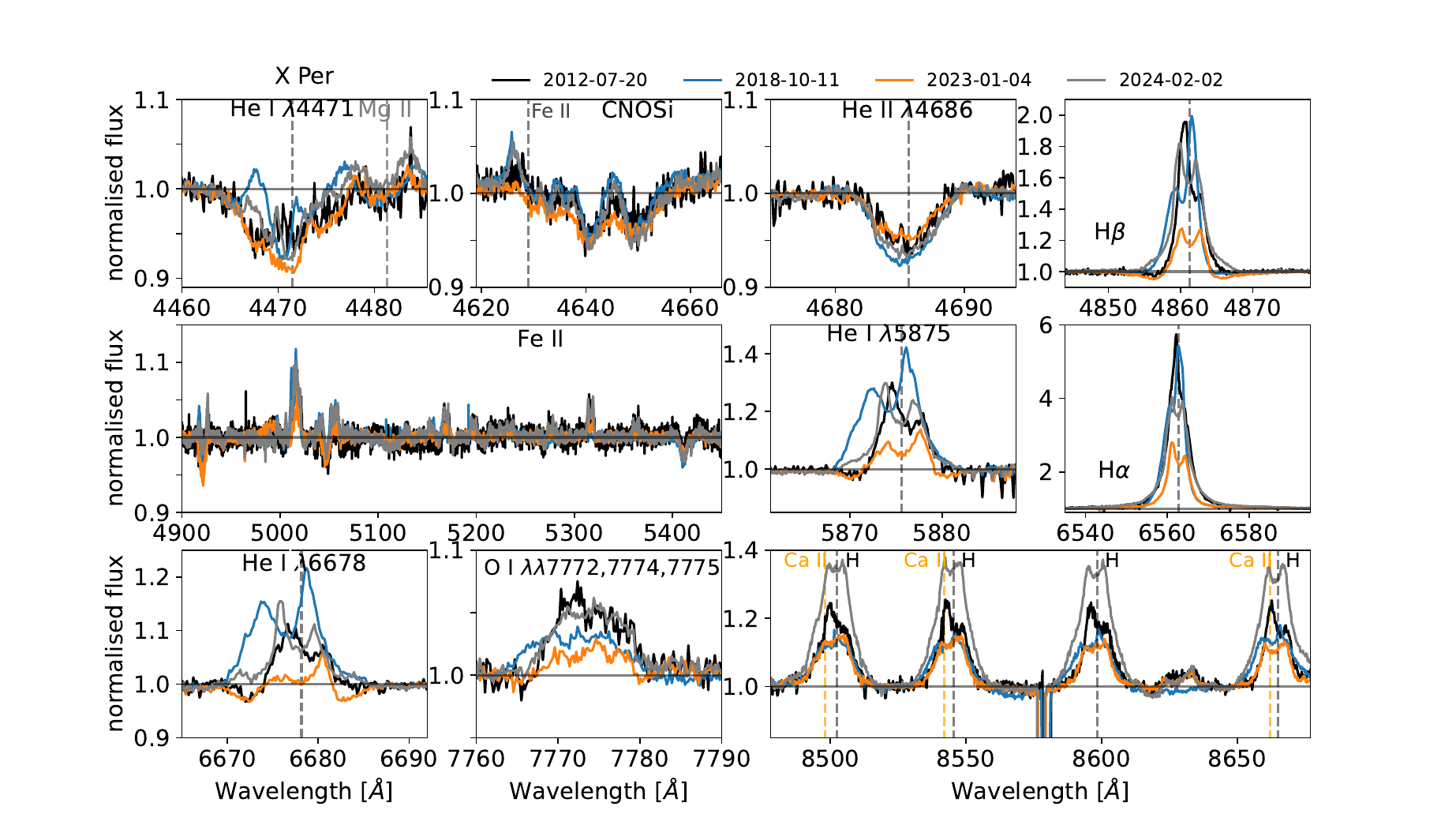}
    \caption{Spectra for \object{X Per} showing strong variability in V/R and line strength.}
    \label{fig_spectral_Xper}
\end{figure*}

\begin{table*}
    \centering
\caption{Nature of typical spectral lines for the different targets. }
    \begin{tabular}{ c|ccccccc }
     \hline
     \hline
spectral line&\object{EM* VES 826}&\object{HD 259440}&\object{V420 Aur}&\object{V615 Cas}&\object{V725 Tau}&\object{V831 Cas}&\object{X Per} \\
\hline
H$\alpha$&dp&e&a&dp&e/dp&dp&e/dp\\
H$\beta$&dp&a + dp&dp&dp&dp&dp/Pc&dp\\
H$\gamma$&a + dp&a + dp&a + dp&a + e&a + dp&a + Pc&a + dp\\
H Paschen ($\lambda$8000-9000)&dp&dp + Ca&dp + Ca&a/dp&dp&a&e\\
\hline
He {\sc i} $\lambda$4009&/&a&a&a&a&a&a\\
He {\sc i} $\lambda$4026&a (weak)&a&a&a&a&a&a\\
He {\sc i} $\lambda$4121&a&a&a&a&a&a&a\\
He {\sc i} $\lambda$4144&a (weak)&a&a&a&a&a&a\\
He {\sc i} $\lambda$4388&a&a + e&a + e&a&a&a&a\\
He {\sc i} $\lambda$4471&a&a&a&a&a + e&a&a + e\\
He {\sc i} $\lambda$4713&a&a (weak)&a&a&a&a + e&dp?\\
He {\sc i} $\lambda$4922&a + Fe?&a + Fe&a + e + Fe&a&a + Fe&a + e&a + e\\
He {\sc i} $\lambda$5016&e&a + Fe&e + Fe&a&dp&dp&dp + Fe\\
He {\sc i} $\lambda$5048&a&a&a&a + Fe?&a + Fe?&a&a + e\\
He {\sc i} $\lambda$5876&dp&a + dp&dp&dp&dp&dp/Pc&dp\\
He {\sc i} $\lambda$6678&dp&a + dp&a + e&dp&dp&Pc&dp\\
\hline
He {\sc ii} $\lambda$4686&a&/&a (weak)&/&a&/&a\\
He {\sc ii} $\lambda$5412&a&/&/&a&a&/&a\\
\hline
O {\sc i} $\lambda\lambda$7772,7774,7775&e (weak)&e&e&a&e&a&e\\
\hline
C-N-O-Si $\lambda\lambda$4630-4655$^{*}$&a&a + Fe&a + Fe&a&a&a&a + e?\\
\hline
Fe {\sc ii} emission lines&/&dp&dp&e (weak)&/&/&e\\
     \hline
    \end{tabular}
         \flushleft
    \begin{tablenotes}
      \small \textbf{Notes.} Explanation of the abbreviations: a = absorption, dp = double-peaked, e = emission non-dp, Fe = Fe {\sc ii} emission, Ca = Ca {\sc ii} emission, Pc = P-cygni like profile. A lone dash indicates absence of the line or a non-detection. A dash in between abbreviations means a changing profile. $^{*}$ Includes the following lines: N {\sc ii} $\lambda\lambda$4631,4643, Si {\sc iv} $\lambda\lambda$4631,4654, N {\sc iii} $\lambda\lambda$4634,4641, O {\sc ii} $\lambda\lambda$4639,4642,4649,4651, and C {\sc iii} $\lambda\lambda$4647,4650,4651.
    \end{tablenotes}
\label{table_spectral_lines_alltargets}
\end{table*}

\subsection{X Per}\label{appendix_spectral_Xper}
\object{X Per} (\object{HD 24534}, \object{4U 0352+30}) is not listed in BESS. It is a persistent X-ray binary with a B0/1\,Ve luminous component \citep{Lyubimkov_et_al_1997,Zamanov_et_al_2019} and an X-ray pulse period of 837\,s \citep[e.g.][]{White_et_al_1976,Haberl_et_al_1998}. From a pulse-time analysis, \citet{Delgado-Marti_et_al_2001} derived an orbital period of 250.3$\pm$0.6\,d and an eccentricity of 0.11. \citet{Yatabe_et_al_2018} estimated a NS mass of around 2$M_{\odot}$ from a fitting of the relation between the period change and the bolometric luminosity of the X-ray emitting companion.\\
\\
For \object{X Per}, the first three spectra are taken between 2012 and 2014, with almost one year apart. Then, eight spectra were taken on consecutive nights in October 2018. The last nine spectra were taken between August 2022 and March 2024.\\
The spectra of \object{X Per} are very variable, especially in V/R variability and line strength. Almost all He {\sc i} lines have emission line contamination. The He {\sc ii} are clearly present in absorption. The CNOSi blend seems to be either not fully blended or to have an emission core. There are also Fe {\sc i} emission lines present. There seems to be no presence of Ca {\sc ii} $\lambda\lambda$8498,8542,8662 emission lines. Spectra are shown in Fig. \ref{fig_spectral_Xper}.\\
From a visual inspection of the spectra, the emission lines seem to be very variable in the blue wing, but much less so in the red wing. This effect seems less prominent in H$\alpha$, but is strongly present in weaker lines, such as the He {\sc i} lines. This will make it difficult to distinguish between disc variability and orbital motion.

\renewcommand{\thefigure}{B\arabic{figure}}
\setcounter{figure}{0}
\renewcommand{\thetable}{B\arabic{table}}
\setcounter{table}{0}
\newpage
\section{The choice of the boundaries for the bisector method}\label{appendix_boundaries_methods}
For the bisector method, we exclude the bottom and top part of the spectral lines. Here, we investigate how the measured RVs depend on the chosen fraction of the line to exclude by varying the percentage of the line that is excluded from the bottom and the top. A part of the top is preferentially excluded because of the large variability that is sometimes seen in the top regions of the emission lines due to variability in the disc.\\
Fig. \ref{fig_appendix_RVs_V420Aur_testcase_bisector} shows RVs measured for \object{V420 Aur} using different cuts on H$\alpha$. The top panel shows the absolute RVs. While the systemic velocities differ between different cuts, the different sets of RVs follow the same general trend. This can be more clearly 
\begin{Figure}
\centering
 \includegraphics[width=\linewidth]{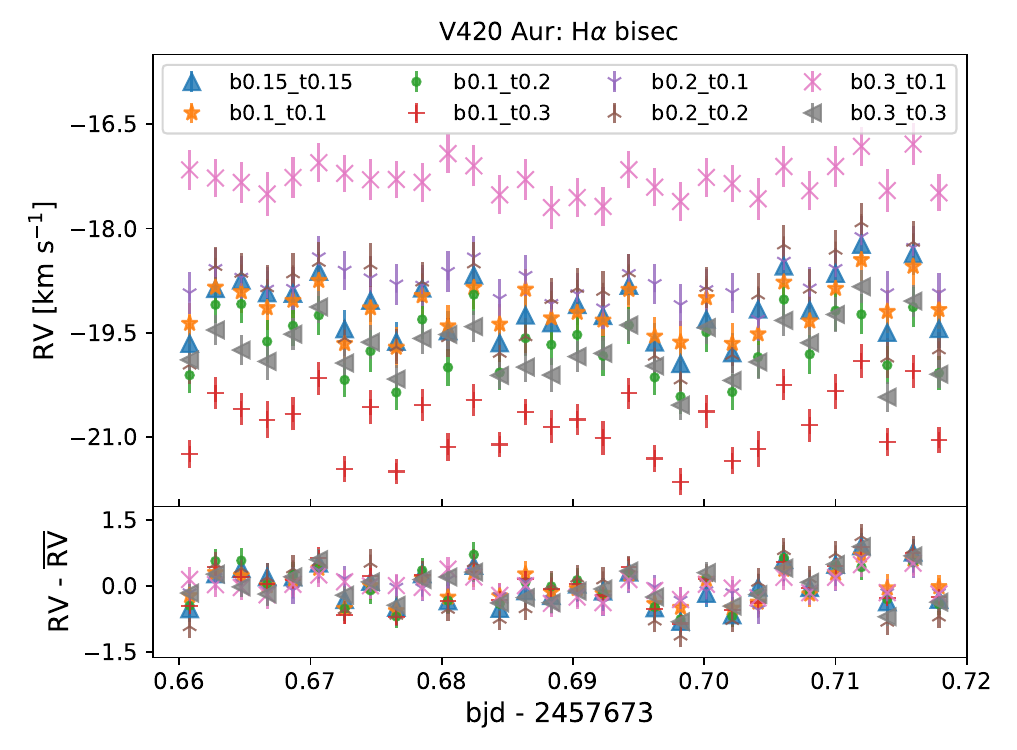}
 \captionof{figure}{Radial velocities for \object{V420 Aur} obtained with the bisector method using different cuts on the H$\alpha$ emission line. Top panel shows absolute RVs. Bottom panel shows the mean-subtracted RVs. In the legend, the fractions that are excluded starting from the bottom and top of the line come after the letter `b' and `t', respectively.}
 \label{fig_appendix_RVs_V420Aur_testcase_bisector}
\end{Figure}
\noindent seen in the mean-subtracted RVs, which are shown in the bottom panel of Fig. \ref{fig_appendix_RVs_V420Aur_testcase_bisector}.\\
The systemic velocities between different cuts are slightly offset from one another, with the largest offsets observed when excluding a larger fraction starting from the top (red pluses) or the bottom (pink crosses). The different systemic velocities indicate that the upper part of the line profile of H$\alpha$ for \object{V420 Aur} is more redshifted and the bottom part more blueshifted. Something similar is also seen when repeating this for H$\beta$. While this reveals how the line is shaped, it is beyond the scope of this work to research in detail the shape of the line profiles.\\
In this work, we are not aiming to determine the systemic velocities of the BeXRBs but determine whether we can see orbital motion. Since the different cuts follow the same general trend, we will further use only one cut. We will choose to exclude 10\% of the bottom and top (orange stars in Fig. \ref{fig_appendix_RVs_V420Aur_testcase_bisector}).

\section{Radial-velocity precision}
Table \ref{table_chi2values} lists the average RV measurement uncertainty $\bar{\sigma}$ and rms values for different spectral lines and methods for a short (V420 Aur:
night of 16 October 2016; HD 259440 and V615 Cas: average value of nights with three consecutive measurements) data set
($\mathrm{rms}_\mathrm{short}$). The results are illustrated in Fig. \ref{fig_RVs_line_precision}.
\end{multicols}

\renewcommand{\thefigure}{C\arabic{figure}}
\setcounter{figure}{0}
\renewcommand{\thetable}{C\arabic{table}}
\setcounter{table}{0}
\begin{table*}[hb]
    \centering
\caption{Values for $\bar{\sigma}$ and $\mathrm{rms}_\mathrm{short}$.}
    \begin{tabular}{ cl|cc|cc|cc}
     \hline
     \hline
&&\multicolumn{2}{c|}{\object{V420 Aur}}&\multicolumn{2}{c|}{\object{HD 259440}}&\multicolumn{2}{c}{\object{V615 Cas}}\\
Method&spectral line&$\bar{\sigma}$\,[km\,s$^{-1}$]&$\mathrm{rms}_\mathrm{short}$\,[km\,s$^{-1}$]&$\bar{\sigma}$\,[km\,s$^{-1}$]&$\mathrm{rms}_\mathrm{short}$\,[km\,s$^{-1}$]&$\bar{\sigma}$\,[km\,s$^{-1}$]&$\mathrm{rms}_\mathrm{short}$\,[km\,s$^{-1}$]\\
\hline
CC&H$\alpha$&0.30&0.25&0.17&0.12&0.48&0.34\\
&H$\alpha$ cut&0.38&0.72&0.19&0.05&0.77&0.48\\
&H$\alpha$ abs&-&-&-&-&1.51&0.28\\
&H$\beta$&0.50&0.71&0.33&0.44&1.07&1.29\\
&H$\beta$ cut&1.25&2.27&0.63&0.47&-&-\\
&H$\beta$ abs&-&-&-&-&1.23&1.13\\
&H$\gamma$&-&-&0.71&0.85&2.08&6.25\\
&H pas&0.66&1.13&0.40&0.49&-&-\\
&Fe&0.88&1.66&0.55&1.00&-&-\\
&He {\sc i} abs&1.63&3.28&3.63&9.11&1.70&6.46\\
&He {\sc i} em&-&-&0.83&1.34&0.70&2.07\\
&O {\sc i} 7774&2.46&4.39&1.29&1.94&-&-\\
&He {\sc ii}&-&-&-&-&4.69&16.05\\
&He {\sc i} + {\sc ii} abs&-&-&-&-&1.61&7.19\\
&He {\sc i} + {\sc ii} + O {\sc i}&-&-&-&-&1.58&6.26\\
&CNOSi&2.22&3.85&2.99&6.77&8.11&39.78\\
&Ca {\sc ii} triplet&-&-&0.34&0.19&-&-\\
\hline
lpf&H$\alpha$&0.23&0.25&0.23&0.14&2.71&3.90\\
&H$\alpha$ double&-&-&-&-&0.62&1.11\\
&H$\alpha$ cut&0.17&0.22&-&-&0.80&0.65\\
&H$\beta$&1.13&1.35&0.32&0.42&2.08${^*}$&1.22${^*}$\\
&H$\beta$ double&0.46&0.78&0.97&4.16&1.21&2.90\\
&H$\beta$ cut&0.52&0.97&-&-&3.51&6.90\\
&H$\gamma$&-&-&-&-&8.69${^*}$&4.20${^*}$\\
\hline
bisec&H$\alpha$&0.24&0.34&0.15&0.16&0.41&1.54\\
&H$\beta$&0.23&1.43&0.08&0.51&0.27&8.03\\
     \hline
    \end{tabular}
         \flushleft
    \begin{tablenotes}
      \small ${^*}$ Gaussian fit was an absorption line.
    \end{tablenotes}
\label{table_chi2values}
\end{table*}

\renewcommand{\thefigure}{D\arabic{figure}}
\setcounter{figure}{0}
\renewcommand{\thetable}{D\arabic{table}}
\setcounter{table}{0}
\begin{multicols}{2}
\section{Periodograms}\label{appendix_extraPs}
\begin{figure*}[hb]
    \centering
    \begin{subfigure}{0.333\linewidth}
    \includegraphics[trim= 9 13 9 13,clip,width = \textwidth]{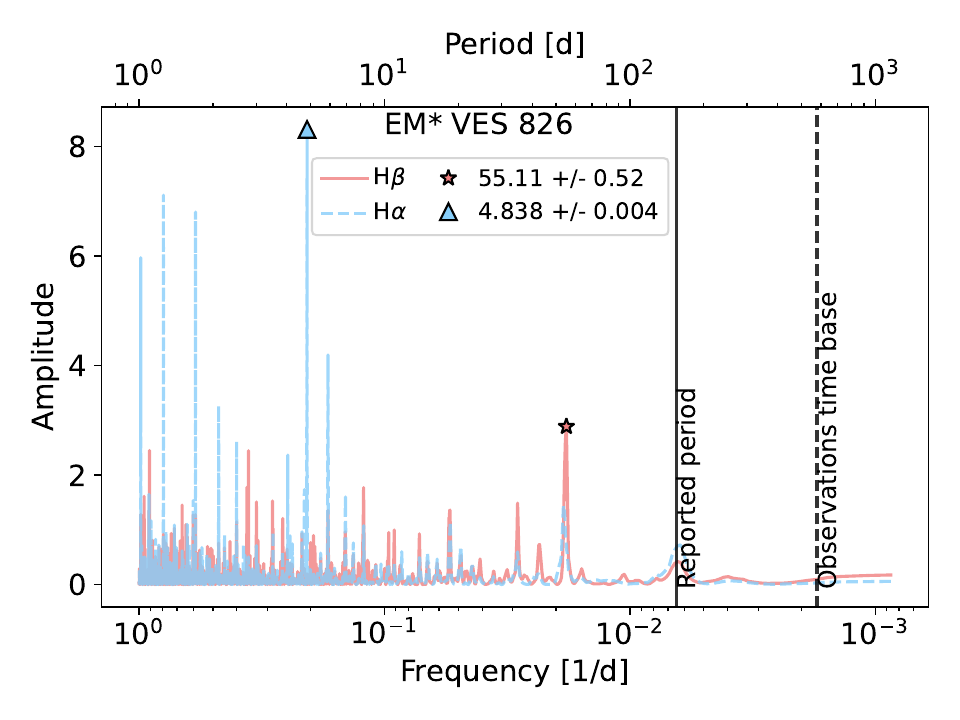}
    \end{subfigure}
    \hspace{-0.1in}
    \begin{subfigure}{0.333\linewidth}
    \includegraphics[trim= 9 13 9 13,clip,width = \textwidth]{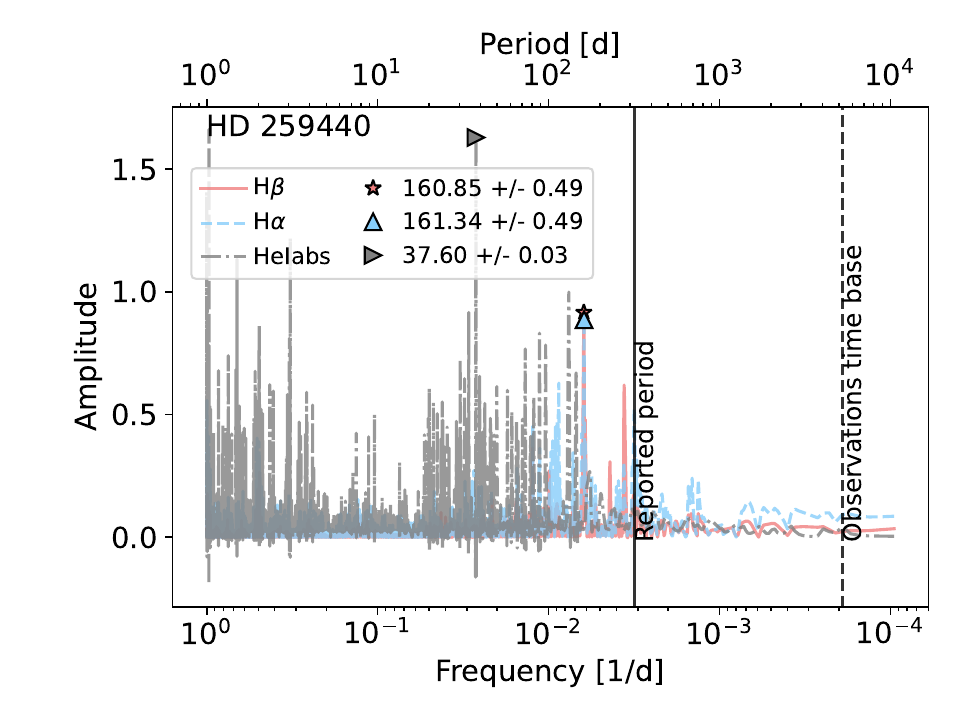}
    \end{subfigure}
    \hspace{-0.1in}
   \begin{subfigure}{0.333\linewidth}
    \includegraphics[trim= 9 12 9 13,clip,width = \textwidth]{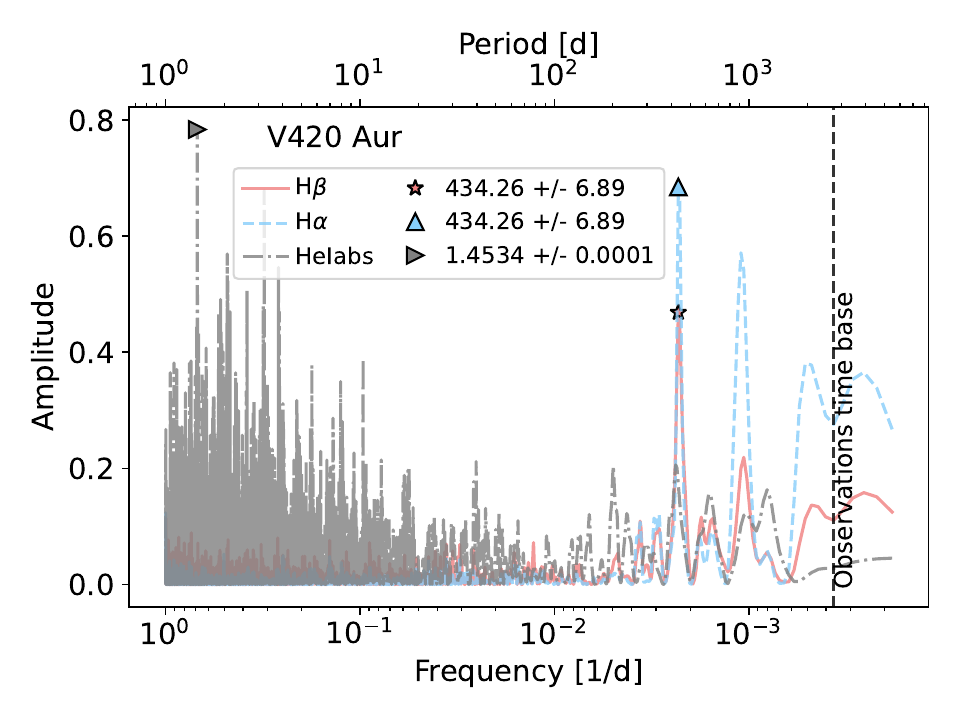}
    \end{subfigure}
    \begin{subfigure}{0.333\linewidth}
    \includegraphics[trim= 9 13 9 13,clip,width = \textwidth]{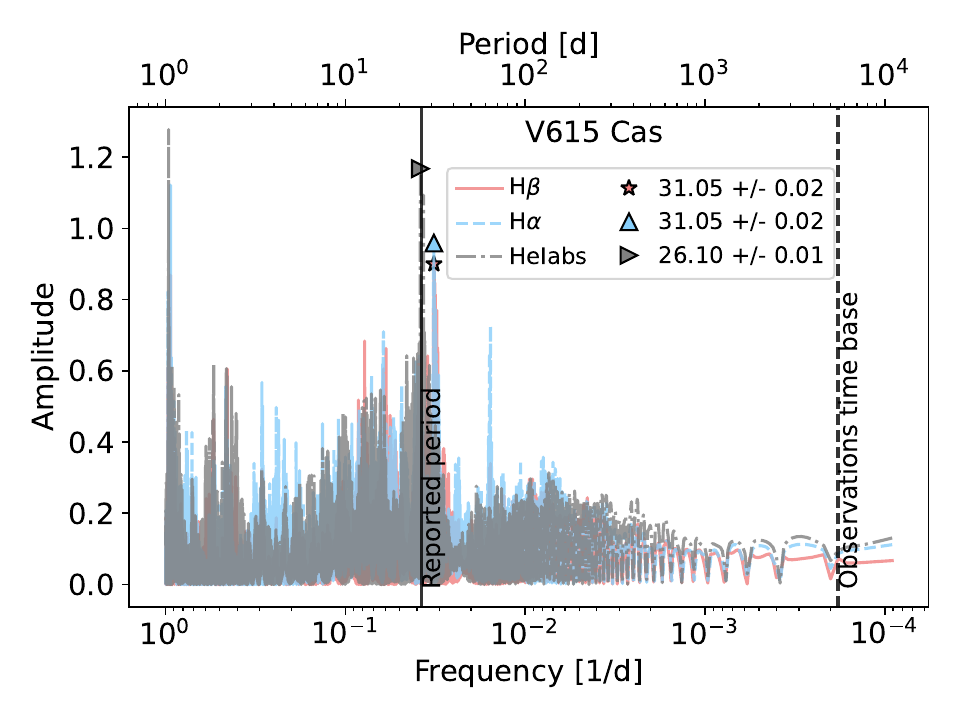}
    \end{subfigure}
    \hspace{-0.1in}
    \begin{subfigure}{0.333\linewidth}
    \includegraphics[trim= 9 13 9 13,clip,width = \textwidth]{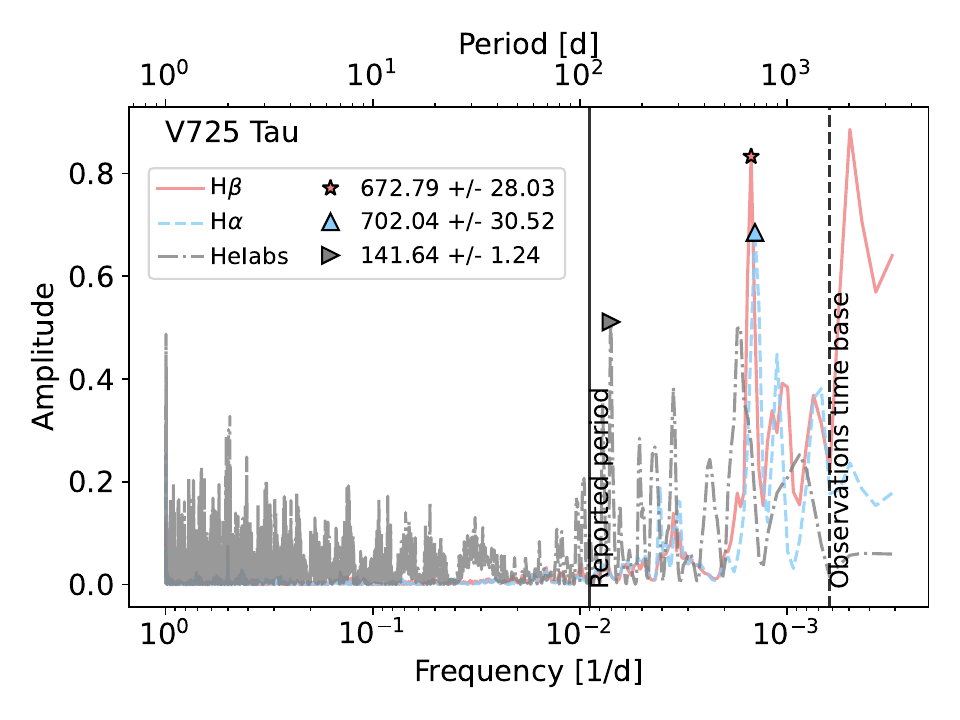}
    \end{subfigure}
    \hspace{-0.1in}
    \begin{subfigure}{0.333\linewidth}
    \includegraphics[trim= 9 13 9 13,clip,width = \textwidth]{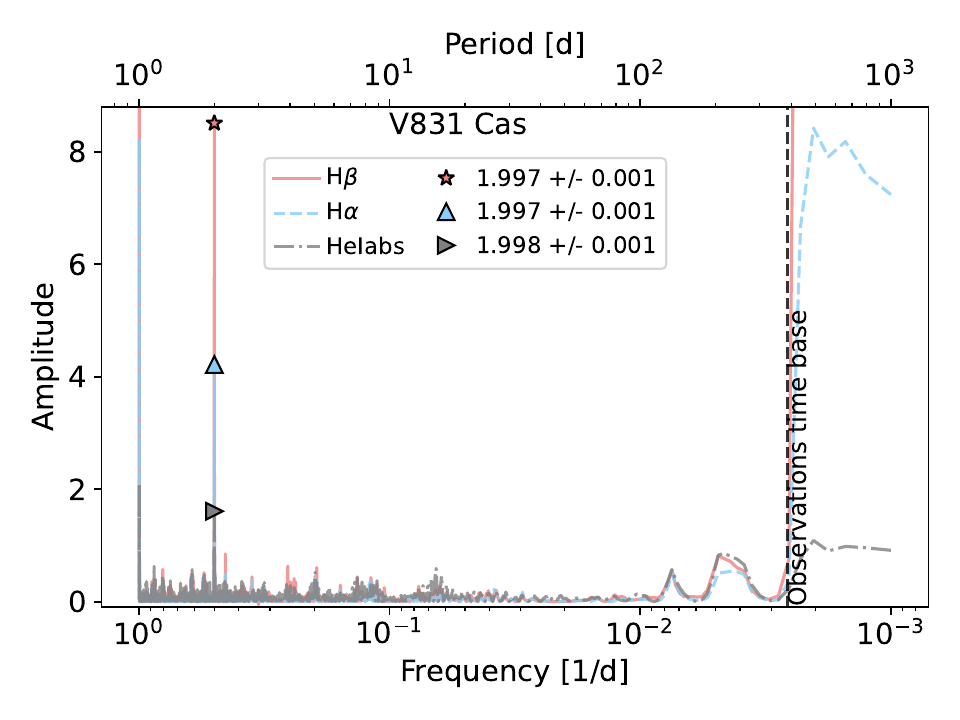}
    \end{subfigure}
    \begin{subfigure}{0.333\linewidth}
    \includegraphics[trim= 9 13 9 13,clip,width = \textwidth]{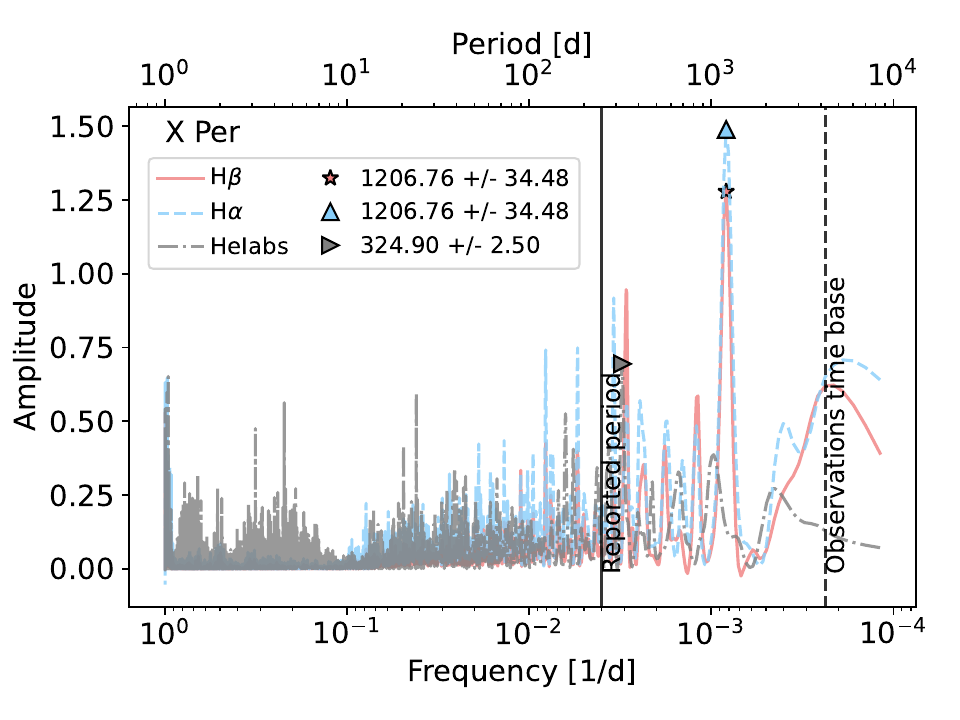}
    \end{subfigure}
    \caption{Periodograms for H$\beta$, H$\alpha$, and He {\sc i} abs (if available). Highest-amplitude frequencies are marked with a black-edged symbol with their value listed in the legends in units of days. Vertical solid-black and dashed-black lines indicate the reported X-ray orbital period and the observation time span, respectively. Each panel shows a different system.}
    \label{fig_periodograms}
\end{figure*}

Here, we present in detail the periodograms of the targets, which were calculated using the metric introduced in \citet{Saha_Vivas_2017}. Instead of using the generalised Lomb-Scargle (gLS) Fourier-transform method \citep{Lomb_1976,Scargle_1982}, \citet{Saha_Vivas_2017} made use of phase-dispersion minimisation \citep[PDM;][]{Stellingwerf_1978}, which is based on the principle that, for the true period, the scatter of the data in different phase bins must be minimal. Moreover, the size of the bins can be varied for different trial periods for sparse data. \citet{Saha_Vivas_2017} showed that the adapted periodogram metric 2gLS/PDM, which was used in this work, is more reliable for sparsely distributed data. Periodograms are shown in Fig. \ref{fig_periodograms}.\\
\\
For none of the seven BeXRBs is the X-ray orbital period the most significant period in the periodograms. This thus already indicates difficulties in finding independent orbital solutions. Hence, for most targets, the information of the X-ray orbital period is crucial to find orbital solutions.\\
One of the main reasons for this is most likely the combination of the sparse sampling of the data for many of the targets and the large spectral changes over the time course of years. Indeed, for \object{EM* VES 826} and \object{HD 259440} there is a local maximum around the reported X-ray period, which could become more significant with more data. \\
\object{V725 Tau} and \object{X Per} are two examples of systems that exhibit large-scale variability (see Figs. \ref{fig_spectral_V725Tau} and \ref{fig_spectral_Xper}). For neither is the X-ray orbital period retrieved. Instead, the periodograms of emission lines suggest a much longer period, likely due to the large blue-shift in the RVs induced by the spectral changes. Even with the absorption lines, the X-ray orbital period is not recovered. Instead, the highest-amplitude period for He {\sc i} abs is longer, but closer to the X-ray orbital period.\\
For \object{V615 Cas}, He {\sc i} abs show a clear period close to the previously reported period of 26.5\,d. For the emission lines, the highest-amplitude period is $\sim$31\,d. Due to the large density of peaks present in the periodograms of \object{V615 Cas}, the X-ray orbital period is not the highest-amplitude period. The 31\,d period might be a beat period due to a precessing disc.

\end{multicols}

\end{appendix}

\end{document}